\documentclass[11pt,paper]{article}

\pdfoutput=1
\usepackage{jheppub}

\usepackage[usenames,dvipsnames,table]{xcolor}
\usepackage{graphicx,amsmath,amssymb,multirow,array,bm,mathrsfs}
\usepackage{epsf,amsfonts}
\usepackage[numbers,sort&compress]{natbib}
\usepackage{dsfont}
\usepackage{graphicx}
\usepackage{slashed}
\usepackage{float}
\usepackage{mathrsfs}
\usepackage{microtype}
\usepackage{mathbbol}

\usepackage{pbox}
\usepackage{amsmath}
\usepackage{amsfonts}
\usepackage{pbox}
\usepackage{multirow}

\usepackage{pict2e}
\makeatletter
\DeclareRobustCommand{\loplus}{\mathbin{\mathpalette\dog@lsemi{+}}}
\newcommand{\dog@rsemi}[2]{\dog@semi{#1}{#2}{-90,90}}
\newcommand{\dog@lsemi}[2]{\dog@semi{#1}{#2}{270,90}}
\newcommand{\dog@semi}[3]{%
  \begingroup
  \sbox\z@{$\m@th#1#2$}%
  \setlength{\unitlength}{\dimexpr\ht\z@+\dp\z@\relax}%
  \makebox[\wd\z@]{\raisebox{-\dp\z@}{%
    \begin{picture}(1,1)
    \linethickness{\variable@rule{#1}}
    \roundcap
    \put(0.5,0.5){\makebox(0,0){\raisebox{\dp\z@}{$\m@th#1#2$}}}
    \put(0.5,0.5){\arc[#3]{0.5}}
    \end{picture}%
  }}%
  \endgroup
}
\newcommand{\variable@rule}[1]{%
  \fontdimen8
  \ifx#1\displaystyle\textfont3\else
    \ifx#1\textstyle\textfont3\else
      \ifx#1\scriptstyle\scriptfont3\else
        \scriptscriptfont3\relax
  \fi\fi\fi
}
\makeatother

\newcommand{\be}{\begin{equation}}
\newcommand{\ee}{\end{equation}}
\newcommand{\bea}{\begin{eqnarray}}
\newcommand{\eea}{\end{eqnarray}}
\newcommand{\tr}{{\text{Tr}}}

\newcommand{\bes}{\begin{equation*}}
\newcommand{\ees}{\end{equation*}}
\newcommand{\beas}{\begin{eqnarray*}}
\newcommand{\eeas}{\end{eqnarray*}}

\newcommand{\vp}{\varphi}

\newcommand{\cM}{\mathcal{M}}

\newcommand{\bmat}{\begin{bmatrix}}
\newcommand{\emat}{\end{bmatrix}}

\def\tr{{\rm tr}}

\def\d{{\rm d}}

\def\cD{{\cal D}}

\def\cH{{\cal H}}

\def\cL{{\cal L}}
\def\cM{{\cal M}}
\def\cN{{\cal N}}
\def\cO{{\cal O}}
\def\cP{{\cal P}}

\newcommand{\ben}{\begin{enumerate}}
\newcommand{\een}{\end{enumerate}}

\newcommand{\Ad}{{\rm Ad}}
\newcommand{\ad}{{\rm ad}}

\title{Geometric Actions for Lower-Spin Gravity}

\affiliation[a]{Physique Math\'{e}matique des Interactions Fondamentales, Universit\'{e} Libre de Bruxelles, and International Solvay Institutes, Campus Plaine - CP 231, 1050 Bruxelles, Belgium}

\author[a]{St\'{e}phane Detournay}
\emailAdd{sdetourn@ulb.ac.be}
\emailAdd{qvdmiers@ulb.ac.be}

\author[a]{and Quentin Vandermiers}

\abstract{

We perform the Hamiltonian reduction of Lower Spin Gravity, the simplest bulk dual for a Warped Conformal Field Theory (WCFT), consisting in an $SL(2) \times U(1)$ Chern-Simons model. We identify the boundary action as the geometric action on coadjoint orbits of the Warped Virasoro group.
We use this reduced action to compute one-loop contributions to the torus partition function and compare them to the Warped Virasoro characters.

}
\begin{document}
\noindent

\maketitle

\section{Introduction and Summary}


Warped Conformal Field Theories (WCFTs) are a class of two-dimensional field theories characterized by chiral scaling symmetries, acting only on right-movers, differing from traditional 2d CFTs \cite{Hofman:2011zj, Detournay:2012pc}. These theories are governed by a Virasoro algebra coupled with a U(1) current algebra, leading to modular covariance of the partition function and providing powerful constraints akin to those in conformal field theories. The studies of free model partition functions \cite{Castro:2015uaa}, correlators \cite{Song:2017czq}, the bootstrap program \cite{Apolo:2018eky}, entanglement entropy \cite{Castro:2015csg, Apolo:2020qjm}, the complex SYK model \cite{Chaturvedi:2018uov}, chaos \cite{Apolo:2018oqv}, anomalies\cite{Jensen:2017tnb} and complexity \cite{Bhattacharyya:2022ren} to name a few have provided deeper insights into their structure and potential physical implications. 

One significant application of WCFTs lies in holography, particularly in describing the near-horizon geometries of extremal black holes \cite{Aggarwal:2019iay}, or so-called Warped AdS$_3$ (WAdS$_3$) black holes \cite{qiao2008edgestatesdestroyeddisordered} whose isometries match the global $SL(2,R) \times U(1)$ subgroup of WCFTs. In the latter case, the asymptotic symmetries are seen to match the full local WCFT symmetries \cite{Masina:2008zv, Compere:2009zj}, the same way the asymptotic symmetries of AdS$_3$ spaces match those of 2d CFTs \cite{Brown:1986nw} (a similar statement holds for 3d flat space, whose asymptotic symmetries match those of a 2d Carrollian CFT \cite{Barnich:2006av}). WCFTs are noted for their non-local Lorentz-violating nature, expanding the scope of holography beyond asymptotically AdS spacetimes. These theories' unique symmetries facilitate the derivation of the Bekenstein-Hawking entropy of black holes, linking quantum gravity insights with field theory descriptions, in the same spirit as \cite{Strominger:1997eq} for BTZ black holes \cite{Banados:1992wn, Banados:1992gq}.

Bulk theories potentially dual to WCFTs typically involve higher curvature terms (such as Topologically Massive Gravity \cite{DESER2000409}) or matter fields \cite{Banados:2005da, Compere:2007in, Israel:2004vv}, and hence degrees of freedom. One notable exception consists in an $SL(2,R) \times U(1)$ Chern-Simons theory, dubbed {\emph Lower-Spin Gravity} introduced in \cite{Hofman:2014loa}. It was argued that this model was the minimal setup for a holographic description of WCFTs, in analogy with the gauge theory formulations of pure gravity with a negative or without cosmological constant\cite{Witten:1988hc} based on Chern-Simons models with $SO(2,2)$ or $ISO(2,1)$ gauge groups respectively. The exploration of the boundary properties of these models started in the AdS$_3$ case with Coussaert, Henneaux, and van Driel who performed a Hamiltonian reduction of the corresponding action. They found that the action can be presented as two chiral Wess-Zumino-Witten (WZW) models, that combine into a single non-chiral WZW. Under the Brown-Henneaux boundary conditions, this action reduces further to the Liouville action on the asymptotic boundary. 
To incorporate BTZ black holes and their descendants, more complex topologies than the filled cylinder of global AdS$_3$ must be considered. This involves using Chern-Simons theory on a disk with a puncture, where the holonomy measures the black hole mass and angular momentum. Repeating the Hamiltonian reduction in this context results in two chiral bosons instead of a combined WZW model, with the zero modes of the bosons related to the bulk holonomy \cite[p. 56--59]{Henneaux:2019sjx}. An equivalent formulation of this action -- named Alekseev-Shatashvili (AS) or geometric action -- arises from the Kirillov-Konstant coadjoint orbit method for the Virasoro group \cite[p. 60]{Henneaux:2019sjx}. The latter could be obtained as a Drinfeld-Sokolov reduction of the SL(2,R) WZW model, which is the CFT counterpart of picking Brown-Henneaux boundary conditions. The connection between pure 3d gravity and coadjoint orbits of the Virasoro group can also be understood as follows. The Bañados geometries \cite[p. 69]{Henneaux:2019sjx} -- the classical saddles of pure 3d gravity with Brown-Henneaux boundary conditions in Fefferman-Graham gauge -- are parametrized by two functions that are the dual stress-tensor expectation values semi-classically. Since the stress-tensor transforms in the coadjoint representation of the Virasoro group \cite[p. 61]{Henneaux:2019sjx}, the Ba˜nados geometries are intrinsically related to the coadjoint orbits of the Virasoro group \cite[p. 57, 70--72]{Henneaux:2019sjx} with the orbit representative corresponding to the global charges of the bulk Banados geometry. Similar ideas can be developed in flat spacetimes, with the Virasoro group being replaced by BMS$_3$ \cite{Merbis:2019wgk}. Under Barnich and Compère's boundary conditions, the Chern-Simons reduces to a two-dimensional boundary theory akin to a flat space version of Liouville theory for vanishing bulk holonomies. This boundary theory is found to coïncide with the geometric action on the coadjoint orbit of the BMS$_3$ group, where orbit representatives correspond to gravitational saddle parameters like mass and angular momentum. The global Minkowski vacuum corresponds to the first exceptional BMS$_3$ orbit, while flat space cosmologies \cite{Cornalba:2002fi, Cornalba:2002nv} -- the flat counterparts of the BTZ black holes --  correspond to generic massive orbits, with orbit representatives corresponding to gravitational charges' zero modes.

The goal of this note will be to study the Hamiltonian reduction of Lower-Spin Gravity under the boundary conditions proposed in \cite{Azeyanagi:2018har}. They were shown the include the "warped counterpart" of the Banados geometries, parametrized by two functions, whose zero modes can be identified with WAdS$_3$ black holes. We will revisit the theory of coadjoint orbits of the Warped Virasoro group and relate the orbit representatives to the Chern-Simons charges of the bulk solutions. We will derive the geometric action on the coadjoint orbits of the warped Virasoro group with generic (constant) orbit representatives, which generalizes the warped Schwarzian action (4.5) of \cite{Afshar:2019tvp} to a generic orbit. Next, we will perform the reduction of Chern-Simons Lower-Spin Gravity with constant charges and observe it matches with the geometric action. As a next step, we compute the one-loop torus partition function of Warped Virasoro descendants around the global vacuum and the Chern-Simons counterparts of WAdS$_3$ black holes. A similar computation was done for AdS$_3$ \cite{Cotler:2018zff} and Minkowski \cite{Merbis:2019wgk} and the results shown to match with the appropriate characters of the corresponding symmetry groups. In this case, the matching is not immediate, and we trace this to the analytic continuation procedure (WCFTs being intrinsically Lorentzian) and to an ambiguity in the definition of WCFT partition functions ("canonical" vs. "quadratic" ensemble). Using the language of coadjoint orbits and geometric actions would in principle allow one to very efficiently compute a variety of interesting physical quantities, such as the Warped conformal blocks, along the lines of \cite{Cotler:2018zff, Merbis:2019wgk}. We leave this as an interesting direction to pursue for future work, alongside a thorough analysis of Lower-Spin Gravity in quadratic ensemble.

\section{Lower Spin Gravity}

Lower Spin Gravity is one of the simplest warped gravity dual, alongside with AdS$_3$ spacetimes with CSS boundary conditions \cite{Compere:2013bya}. It consists in an $SL(2, \mathbb R) \times U(1)$ Chern-Simons action
 \begin{equation} \label{LowerSpinAction}
     S = \frac{k}{4 \pi} \int_{\cM}  \left< A \wedge \d A + \frac{2}{3} A \wedge A \wedge A \right> +\frac{\kappa}{2\pi} \int_{\cM} \left< C \wedge \d C \right> \, ,
 \end{equation}
 where $k$ is a continuous parameter determining the central charge of the Virasoro algebra, and $\kappa$ a discrete parameter describing the sign of the level of the $U(1)$ Kac-Moody algebra. It was first constructed in \cite{Hofman:2014loa} from the coupling of WCFTs to a background geometry. 
 
A consistent set of boundary conditions on the connections $A$ and $C$ for this model was proposed in \cite{Azeyanagi:2018har}: 
 
\begin{align} \label{BC_LSG}
\begin{split}
    A_\vp = L_1 - \mathfrak L \, L_{-1} \quad &, \quad A_t = \mu \, A_\vp - \mu' \, L_0 + \frac{\mu''}{2} L_{-1} \, , \\
    C_\vp = \frac{2\pi}{\kappa} \cP \quad &, \quad C_t = \mu \, C_\vp + \nu \, , 
    \end{split}
\end{align}
where
\begin{equation}
    \mathfrak L = \frac{2\pi}{k} \left( \cL - \frac{ \pi}{\kappa} \cP^2 \right) \, ,
\end{equation}
$L_i$ are $SL(2, \mathbb R)$ generators, $(t,\vp)$ the boundary coordinates, and the prime denotes a derivative with respect to $\vp$.
The functions $\cL$ and $\cP$ appearing in the boundary conditions depend on $t$ and $\vp$ and we can interpret them as functions characterizing the physical state. Hence, $\mu$ and $\nu$, also functions of $t$ and $\vp$, can be viewed as chemical potentials and are therefore fixed, i.e. $\delta \mu = 0 = \delta \nu$.
Those boundary conditions were shown to be preserved by warped conformal transformations represented by the following charge algebra:
\begin{align} \label{WCFT algebra}
    \begin{split}
        [L_n, L_m] &= (n-m) \, L_{n+m} + \frac{c}{12} \, n^3 \, \delta_{n+m} \ , \\
        [L_n, P_m] &= - m \, P_{n+m} \, , \\
        [P_n, P_m] &= \kappa \, n \, \delta_{n+m} \, ,
    \end{split}
\end{align}
with $c=6k$. We recognize the Virasoro-Kac-Moody algebra of a WCFT. One can perform a metric interpretation of the connection and show that the boundary conditions contain the different warped AdS$_3$ spacetimes \cite{Hofman:2014loa}, as well as the warped black hole in the canonical ensemble \cite{Azeyanagi:2018har}. 

\section{Coadjoint representation}
The warped conformal group is the centrally extended semi-direct product group of diffeomorphisms of the circle ${\rm Diff}^+(S^1)$ with $C^{\infty}(S^1)$. In Appendix A of \cite{Afshar:2015wjm}, the coadjoint representations of this group were studied, which we will summarize here for convenience and to fix conventions.

We take $\vp$ to be a coordinate along the circle and denote elements of ${\rm Diff}^+(S^1)$ by $f(\vp)$, satisfying $f'(\vp)>0 $ and $f(\vp +2\pi) = f(\vp) + 2 \pi$, and elements of $C^{\infty}(S^1)$ by $p(\vp)$. The group multiplication of the semi-direct product group $G = {\rm Diff}^+(S^1) \ltimes C^\infty(S^1)$ is
\begin{equation}
    (f_1, p_1) \cdot (f_2, p_2) = (f_1 \circ f_2, p_1 + \sigma_{f_1} p_2) \,, \qquad \text{with} \; \sigma_f p = p \circ f^{-1}\, 
\end{equation}
where $\circ$ denotes functional composition, $f \circ \vp = f(\vp)$.

Since the second cohomology space of $G$ is three-dimensional, the warped conformal group has three central extensions. They are defined by the three 2-cocycles in $G$. Denoting elements of the centrally extended group $\hat{G} = G \times \mathbb{R}^3$ by $(f,p; \lambda , \mu, \nu )$, the group operation now reads:
\begin{align}
    (f_1, p_1; \lambda_1, \mu_1, \nu_1) \cdot 
    (f_2, p_2; \lambda_2, \mu_2, \nu_2) = & \,
    \big(f_1 \circ f_2 , p_1 + \sigma_{f_1} p_2; \lambda_1 + \lambda_2 + B(f_1, f_2), \nonumber \\
    & \mu_1 + \mu_2 + C(f_1,p_2), \nu_1 + \nu_2 + D(p_1, \sigma_{f_1}p_2) \big) \,.
\end{align}
Here $B,C$ and $D$ are real-valued 2-cocycles in the group. They are given explicitly by
\begin{align}
\begin{split}
    B(f_1, f_2) & = - \frac{1}{48\pi} \oint \d\vp \;  \log( \partial_\vp f_1 \circ f_2 ) \partial_\vp \log \partial_\vp f_2 \,,  \\
    C(f_1, p_2) & = - \frac{1}{2\pi} \oint \d\vp \; p_2 \partial_\vp \log( \partial_\vp f_1) \,, \\
    D(f_1, \sigma_{f_1} p_2) & = - \frac{1}{4\pi} \oint \d\vp \; p_1 \partial_\vp \sigma_{f_1} p_2 \,.
\end{split}
\end{align}
We recognize for the 2-cocycle $B(f_1, f_2)$, the Bott-Thurston cocycle of the Virasoro group \cite{Unterberger:2012, Guieu:2007}.

The adjoint action of a group element $g_1 = (f_1,p_1;\lambda_1, \mu_1, \nu_1) $ on an element of the algebra $(X_2, p_2; \lambda_2, \mu_2, \nu_2)$ is computed as
\begin{equation}
\Ad_{g_1} (X_2, p_2; \lambda_2, \mu_2, \nu_2) = \frac{\d}{\d\epsilon} \left. g_1 \cdot (e^{\epsilon X_2}, \epsilon p_2, \epsilon \lambda_2, \epsilon \mu_2, \epsilon \nu_2 ) \cdot g_1^{-1} \right|_{\epsilon = 0} \,,
\end{equation}
which explicitly evaluates to
\begin{align}\label{adjoint}
\Ad_{(f,p_1)} (X, p_2; \lambda, \mu, \nu) = & \; \Big( \Ad_{f} X, \sigma_{f} p_2 + \Sigma_{\Ad_{f} X} p_1 ; \lambda - \frac{1}{24\pi} \oint \d\vp \; X(\vp) \{f; \vp \} , \nonumber \\
& \nonumber \mu -\frac{1}{2\pi} \oint \d\vp \; \left( p_2 \partial_\vp \log f' - X(\vp) f'{}^2 \partial_f^2 ( p_1 \circ f) \right) , \\
& \nu - \frac{1}{2\pi} \oint \d\vp \; \left( p_1 \partial_\vp \sigma_f p_2 - \frac12 X(\vp) [\partial_\vp( p_1 \circ f)]^2 \right) \Big) \,.
\end{align}
Here, the notation $\Ad_f X$ denotes the adjoint action of the Virasoro group on the components of a vector field $X(\vp)\partial_\vp$, i.e. the transformation of a vector field on the circle under diffeomorphisms: $\Ad_f X = f'(f^{-1}) X(f^{-1})$.

The symbol $\Sigma_X$ denotes the infinitesimal version of $\sigma_f$, and hence $\Sigma_X p_1 = X(\vp) p_1'(\vp) $. Finally, $\{f;\vp\}$ denotes the Schwarzian derivative
\begin{equation} \label{Scharzian derivative}
    \{f ; x \} = \frac{f^{\prime\prime\prime}}{f'} - \frac{3}{2} \left(\frac{f^{\prime \prime}}{f'}\right)^2 \, .
\end{equation}

The adjoint representation of the algebra on itself gives the commutators
\begin{align} \label{algebra}
- \ad_{(X_1, p_1)} \left( X_2, p_2; \lambda_2, \mu_2, \nu_2 \right) & = [ (X_1, p_1; \lambda_1, \mu_1, \nu_1), (X_2, p_2; \lambda_2, \mu_2, \nu_2) ] \\
& \nonumber = \Big( X_1 X_2' - X_1' X_2, X_1 p_2' - p_1' X_2 ; \frac{1}{24\pi} \oint \d\vp \; X_1''' X_2 , \\
& \nonumber \qquad \frac{1}{2\pi} \oint \d\vp \; \left( X_1'' p_2 - p_1'' X_2 \right), \frac{1}{2\pi} \oint \d\vp\; p_1 p_2' \Big)\,.
\end{align}
From here we can read off the warped conformal algebra, by taking the generators to be $L_n = (e^{ i n \vp} , 0 ; 0, 0,0 ), P_n = (0 , e^{ i n \vp} ; 0, 0,0 )$ and $Z_1 = (0, 0; 1, 0 ,0), Z_2 = (0, 0; 0, 1, 0) $ and $Z_3 = (0, 0; 0, 0 ,1)$:
\begin{align}
\begin{split}
i [L_n, L_m] & = (n-m) L_{n+m} + \frac{Z_1}{12} n^3 \delta_{n+m} \,, \\
i [L_n, P_m] & = - m P_{m+n} - i n^2 Z_2 \delta_{n+m} \,, \\
i [P_n, P_m] & = - m Z_3 \delta_{n+m} \,.
\end{split}
\end{align}

Elements of the dual space $\mathfrak{g}^*$ will be denoted by $(\cL, \cP; c, k, \kappa)$. We denote by $\langle , \rangle $ the pairing between the dual space and the algebra, i.e. $\langle, \rangle $ is a map from $\mathfrak{g}^* \times \mathfrak{g} \to \mathbb{R}$. In our case, we will take
\begin{equation}\label{bracket}
	\langle (\cL, \cP ; c, k , \kappa) , (X, p; \lambda, \mu, \nu) \rangle = \oint \d\vp \left( \cL(\vp) X(\vp) + \cP(\vp) p(\vp) \right) + c \lambda + k \mu + \kappa \nu \,.
\end{equation}
We can now define the coadjoint action of the group on elements of the dual space as
\begin{equation}
\langle \Ad_{(f,p_1)^{-1}}^* (\cL, \cP ; c, k , \kappa) , (X, p_2; \lambda, \mu, \nu) \rangle = \langle (\cL, \cP ; c, k , \kappa) , \Ad_{(f,p_1)} (X, p_2; \lambda, \mu, \nu) \rangle \,.
\end{equation}
By using \eqref{adjoint} and \eqref{bracket} we obtain
\begin{equation}
\Ad_{(f,p_1)^{-1}}^* (\cL, \cP ; c, k , \kappa)  =  (\tilde \cL, \tilde \cP ; c, k , \kappa) \,,
\end{equation}
with 
\begin{align}\label{coadjoint}
\tilde \cL (\vp) & = f'{}^2 \left( \cL(f) + \cP(f) \partial_f (p \circ f) - \frac{c}{24\pi} \frac{ \{ f; \vp \}}{f'{}^2} + \frac{k}{2\pi} \partial_f^2 (p \circ f) + \frac{\kappa}{4\pi}  [\partial_f (p \circ f)]^2 \right) \,,  \nonumber \\
\tilde \cP(\vp) & = f' \left( \cP(f) - \frac{k}{2\pi} \frac{f''}{f'^2} + \frac{\kappa}{2\pi} \partial_f (p \circ f) \right) \,.
\end{align}
The notation here differs from the appendix A of \cite{Afshar:2015wjm}, but this is due to the fact that there it is the quantity $\Ad^*_{(f,p)} (\cL, \cP ; c, k , \kappa) (f(\vp)) $ that is computed, whereas here we have given $\Ad^*_{(f,p)^{-1}} (\cL, \cP ; c, k , \kappa) (\vp) $. For completeness, we also list the other expression here, which is equivalent to \eqref{coadjoint}:

\begin{align}
		\Ad_{(f,p)}^* \cL (f(\vp)) & = \frac{1}{f'{}^2} \bigg( \cL(\vp) - \cP(\vp) (p \circ f)'(\vp) + \frac{c}{24\pi}  \{ f; \vp \} - \frac{k}{2\pi} (p \circ f)''(\vp) \nonumber \\
		& \nonumber \qquad  + \frac{\kappa}{4\pi}  [ (p \circ f)'(\vp)]^2 \bigg) \,,  \\
		\Ad_{(f,p)}^*  \cP(f(\vp)) & = \frac{1}{f'} \left( \cP(\vp) + \frac{k}{2\pi} \frac{f''}{f'} - \frac{\kappa}{2\pi} (p \circ f)'(\vp) \right) \,. \nonumber
\end{align}

\section{Coadjoint orbits}
Here we classify the coadjoint orbits of the warped Virasoro group and their corresponding little group. 
The main point is that coadjoint orbits $\cO_{(\cL_0,\cP_0)} $ are defined as all elements of the dual space $(\cL, \cP)$ which can be obtained from a fixed representative $(\cL_0, \cP_0)$ by the coadjoint action \eqref{coadjoint}. The orbit is isomorphic to the symplectic manifold $\cO \sim G/\cH_{(\cL_0,\cP_0)}$, where $G$ is the group manifold and $\cH_{(\cL_0,\cP_0)}$ is the stabilizer subgroup of the orbit, consisting of all elements which leave $(\cL_0, \cP_0)$ invariant under the coadjoint action \eqref{coadjoint}. From now on we will take $(\cL_0,\cP_0)$ to be constant. Through the reduction of the Chern-Simons theory with constant charges, we should be able to relate the orbit representatives $(\cL_0, \cP_0)$ to the Chern-Simons charges of the bulk solution.
Generically, the infinitesimal coadjoint action \eqref{coadjoint} gives
\begin{align}\label{coadjointalg}
\begin{split}
\ad^*_{\epsilon_L, \epsilon_P} \cL(\vp) = & \epsilon_L \cL' + 2 \epsilon_L' \cL - \frac{c}{24\pi} \epsilon_L''' + \epsilon_P' \cP + \frac{k}{2\pi} \epsilon_P''\,, \\
\ad^*_{\epsilon_L, \epsilon_P} \cP(\vp) = & \epsilon_L' \cP + \epsilon_L \cP' - \frac{k}{2\pi} \epsilon_L'' + \frac{\kappa}{2\pi} \epsilon_P' \,.
\end{split}
\end{align}
We will here focus on orbits with constant representatives, hence they as obtained from the coadjoint action of the algebra on constant $\cL_0$ and $\cP_0$, in which case the above formula simplifies to
\begin{align}\label{constantrep}
\ad^*_{\epsilon_L, \epsilon_P} \cL_0 = &  2 \epsilon_L' \cL_0 - \frac{c}{24\pi} \epsilon_L''' + \epsilon_P' \cP_0 + \frac{k}{2\pi} \epsilon_P''\,, \\
\ad^*_{\epsilon_L, \epsilon_P} \cP_0 = & \epsilon_L' \cP_0 - \frac{k}{2\pi} \epsilon_L'' + \frac{\kappa}{2\pi} \epsilon_P' \,. \label{constantrep2}
\end{align}
As we will see below, whenever $\kappa \neq 0$, one can make a field redefinition such that $k = 0$, so we should distinguish two cases, $\kappa = 0$ and $k \neq 0$ or $\kappa \neq 0$ and $k= 0$ addressed below: 





\subsection{\texorpdfstring{$\kappa=0$}{} and \texorpdfstring{$k \neq 0$}{}}


Let us first classify the orbits $\cO_p$ for $p \in \mathfrak{g}^*$ under the $\sigma$-action of $G$. This amounts to solving the differential equation
\begin{equation}
	\epsilon_L' \cP_0 - \frac{k}{2\pi} \epsilon_L''  = 0\,,
\end{equation}
for $\epsilon_L$ a function on the $S^1$. The general solution is
\begin{equation}
	\epsilon_L = a + b\,  e^{ \frac{2\pi \cP_0 }{k} \vp} \,.
\end{equation}
This is only well defined on the $S^1$ whenever $\cP_0 = i k n / 2\pi$ and hence there are two distinct cases
\begin{itemize}
	\item $\cP_0 \neq i k n / 2\pi$: $b =0$ and  the little group $G_p$ is $U(1)$
	\item $\cP_0 = i k n / 2\pi$: $b \neq 0$ and the little group $G_p$ is two-dimensional: $i[L_0,L_{n}] = - n L_{n}$ for $L_{n} = e^{i n \vp}$ and $L_0 = 1$
\end{itemize}
In the first case, the remaining equation \eqref{constantrep} gives 
\begin{equation}
	\epsilon_P' \cP_0 + \frac{k}{2\pi} \epsilon_P'' = 0
\end{equation}
which is solved by $\epsilon_P = c + d e^{-\frac{2\pi \cP_0}{k} \vp}$. This once again gives only one solution on the circle, under the assumption that $\cP_0 \neq i k n / 2\pi$. Hence these orbits are characterized by a two-dimensional Abelian little group.

The second case is more interesting. In that case the remaining equation \eqref{constantrep} gives
\begin{equation}
i n \epsilon_P ' + \epsilon_P'' = - \frac{2 \pi i n}{k} (2 \cL_0 + \frac{c}{24\pi} n^2) e^{i n \vp} \equiv a_n e^{in \vp} \,. 
\end{equation}
In this case, the solution reads
\begin{equation}
\epsilon_P = c + d e^{-i n \vp} - \frac{a_n}{2 n^2} e^{in\vp} \,.
\end{equation}
Hence, {\it regardless} of the value of $\cL_0$, this solution is well defined on the circle, however, only when $a_n = 0$, or, whenever
\begin{equation}
\cL_0 = - \frac{c}{48\pi} n^2 \,,
\end{equation}
does the algebra $[(\epsilon_L, \epsilon_P),(\epsilon_L, \epsilon_P)]$, understood as \eqref{algebra}, close linearly. In that case, the algebra written in the form of the generators $L_0 = (1,0)$, $L_{n} = ( e^{in\vp},0)$, $P_0 = (0,1)$, and $P_{-n} = (0, e^{-in\vp})$ reads:
\begin{align} \label{little group algebra}
    \begin{split}
    i[L_0, L_n] = - n L_n\,&, \qquad i[L_0, P_{-n}] = n P_{-n}\,, \\
    i[L_n, P_{-n}] = n P_0\,&, \qquad i[P_0,L_n] = 0 \, .
    \end{split}
\end{align}
In the case of $n=1$ this gives exactly the $P_2^c$ algebra, or the central extension of the two dimensional Poincar\'e algebra.

To summarize, if $\cP_0 \neq i k n / 2\pi$, the little group is $U(1) \times U(1)$ and for $\cP_0 = i k n / 2\pi$, it necessarily implies that $\cL_0 = - \frac{c}{48\pi} n^2$ and the little group has the algebra \eqref{little group algebra}.


\subsection{\texorpdfstring{$\kappa\neq 0$}{}}

In this case, the analysis in slightly different, because the semi-direct product does not act on an Abelian group. It is always possible to set $k$ to zero when $\kappa \neq 0$ \cite{Afshar:2015wjm}. Indeed, by making the redefinition
\begin{equation}
    \epsilon_P \to \epsilon_P + \frac{k}{\kappa} \epsilon_L' \, ,
\end{equation}
equation \eqref{constantrep2}, set to zero in order to analyze the orbits of the $\sigma$ action of $G$ on $C^{\infty}(S^1)$, gives
\begin{equation} \label{ep in terms of el}
	\epsilon_P = a - \frac{2\pi \cP_0}{\kappa} \epsilon_L \,.
\end{equation} 
Note that now there is no special value for $\cP_0$.
The gauge parameter $\epsilon_P$ is completely fixed in terms of $\epsilon_L$, up to a constant $a$ (which represents the $U(1)$ factor in the little group). 

Plugging the above into \eqref{constantrep}, redefining
\begin{equation}
    c_{eff} = c - 12 \frac{k^2}{\kappa} \, ,
\end{equation}
and setting this to zero gives
\begin{equation}
 \epsilon_L' \mathfrak{L_0} - 4 \epsilon_L''' = 0\,,
\end{equation} 
where
\begin{equation}
    \mathfrak{L}_0 = \frac{12 \pi}{c_{eff}} \left( \cL_0 - \frac{\pi \cP_0^2}{\kappa} \right) \, .
\end{equation}

This is the usual differential equation which characterizes the little group of Virasoro coadjoint orbits. Hence there are two distinct cases
\begin{itemize}
	\item $\mathfrak{L}_0 \neq - \frac{n^2}{4} $. These are the generic orbits with $U(1)$ little group.
	\item $\mathfrak{L}_0 = - \frac{n^2}{4} $. In this case the orbits' little group is enhanced to the $n$-fold cover of $SL(2,\mathbb{R})$.  
\end{itemize}
Combined with the previous $U(1)$ little group in \eqref{ep in terms of el}, this implies that coadjoint orbits of the $\kappa \neq 0$ warped Virasoro group are characterized by the value of the Sugawara shifted $\mathfrak{L}_0$. For generic values of this quantity there is a two-dimensional Abelian little group, whereas for  $\mathfrak{L}_0 = - \frac{n^2}{4} $ the little group on the coadjoint orbits is $SL^{(n)}(2,\mathbb{R}) \times U(1)$. 

\section{Geometric actions}
The coadjoint orbits form symplectic manifolds, with the symplectic structure given by the Konstant-Kirillov symplectic form $\omega$. For the warped Virasoro group, this has been constructed in \cite{Afshar:2019tvp}, which we follow here by slightly adapting notations:
\begin{equation}\label{omega1}
\omega_{12} =   \langle (\cL, \cP ; c, k , \kappa)  ,\ad_{(X_1, p_1)} \left( X_2, p_2; \lambda_2, \mu_2, \nu_2 \right ) \rangle \,.
\end{equation}
For the generic construction of the geometric actions for centrally extended semi-direct product groups, such as the warped Virasoro group, see also \cite{Barnich:2017jgw}.
To consider the orbit $\cO_{(\cL_0,\cP_0)}$, we take the dual space elements to be  
\begin{align} 
	\cL (\vp) & = f'{}^2 \left( \cL_0 + \cP_0 \partial_f (p \circ f) - \frac{c}{24\pi} \frac{ \{ f; \vp \}}{f'{}^2} + \frac{k}{2\pi} \partial_f^2 (p \circ f) + \frac{\kappa}{4\pi}  [\partial_f (p \circ f)]^2 \right) \,, \nonumber \\
	\cP(\vp) & = f' \left( \cP_0 - \frac{k}{2\pi} \frac{f''}{f'^2} + \frac{\kappa}{2\pi} \partial_f (p \circ f) \right) \,. \label{coadjointL0P0}
\end{align}
Using this, \eqref{omega1} reads explicitly
\begin{align}
\omega_{12} = & \; - \oint \d\vp \; \left( f'^2 (\cL_0 + \partial_f (p \circ f) \cP_0) \left( X_1 X_2' - X_2 X_1'\right) +  \cP_0  f' \left(X_1 p_2' - X_2 p_1' \right) \right) \nonumber \\
& + \frac{c}{24\pi} \oint \d\vp \; \left( X_1'' X_2' + \{f,\vp\} (X_1 X_2' - X_2 X_1') \right)  \\
& + \frac{k}{2\pi} \oint \d\vp \; \left(X_1' p_2' - X_2' p_1' - f'^2 \partial_f^2 (p \circ f) (X_1 X_2' - X_2 X_1') + (\log f') ' (X_1 p_2' - X_2 p_1') \right) \nonumber \\
& - \frac{\kappa}{4\pi} \oint \d\vp \; \left(\mathbf{2} p_1 p_2' + (p \circ f)'{}^2 (X_1 X_2' - X_2 X_1') + 2 (p \circ f)' (X_1 p_2' - X_2 p_1') \right) \nonumber
\end{align}
In terms of differential forms, the above expression can be brought in the form
\begin{align}
\omega = & - \oint \d\vp \; \left(  \cL_0 f'{}^2 \d X \wedge \d X' + \cP_0 f' \d X \wedge (\d p + (p \circ f)' \d X)'  \right) \nonumber\\
& - \frac{c}{48\pi} \oint \d\vp \; \left( \d X' \wedge \d X'' - 2 \{f; \vp \} \d X \wedge \d X'\right) \nonumber \\
& + \frac{k}{2\pi} \oint \d\vp \; \left(\d X' + (\log f')' \d X) \wedge (\d p + (p \circ f)' \d X)' \right)  \\
& - \frac{\kappa}{4\pi} \oint \d\vp \; (\d p + (p \circ f)' \d X) \wedge (\d p + (p \circ f)' \d X)' \nonumber
\end{align}
We now change variables to the finite transformations, by using that 
\begin{equation}
	\d f = f' \d X \,, \qquad  \d g = \d p + \partial_\vp (p \circ f) \d X\,,
\end{equation}
where $g \equiv p \circ f$ does not only transform under ${\rm Diff}(S^1)$ but also under $C^{\infty} (S^1)$.
The result simplifies even further by writing it as
\begin{align}
	\omega = - \oint \d\vp \; \Big(& \cL_0 \d f \wedge \d f' + \cP_0 \d f \wedge \d g' \\
	 &  + \frac{c}{48\pi} \d \log f'  \wedge (\d \log f')' - \frac{k}{2\pi} \d \log f'  \wedge \d g' + \frac{\kappa}{4\pi} \d g \wedge \d g' \Big). \nonumber
\end{align}
It is now strikingly clear that whenever $\kappa \neq 0$, the term proportional to $k$ can be removed by redefining
\begin{equation}\label{redef}
g = \tilde{g} + \frac{k}{\kappa} \log f'\,.
\end{equation}
The result is (dropping a total derivative term proportional to $\cP_0$):
\begin{align}
\omega = - \oint \d\vp \; \Big(& \cL_0 \d f \wedge \d f' + \cP_0 \d f \wedge \d g'  + \frac{c_{\rm eff}}{48\pi} \d \log f'  \wedge (\d \log f')' + \frac{\kappa}{4\pi} \d g \wedge \d g' \Big). \label{KK symp form}
\end{align} 
where 
\begin{equation}
	c_{\rm eff} = c - 12 \frac{k^2}{\kappa}\,.
\end{equation}

Now that we have the Konstant-Kirillov symplectic form for any orbit with constant representatives $(\cL_0, \cP_0)$,  we can obtain the geometric action on this orbit. Since $\d \omega = 0 $, we can locally write $\omega = \d \alpha$. Then the kinetic part of the geometric action on the coadjoint orbit is the integral over the orbit of $\alpha$, i.e.
\begin{align}
I[f,g;\cL_0,\cP_0] & = \int_{\gamma} \alpha \\
& = \int_\gamma \oint \d\vp \Big( \cL_0 f' \d f + \cP_0 f' \d g + \frac{c}{48\pi} \frac{f'' \d f'}{f'{}^2} + \frac{k}{2\pi} \frac{f''}{f'} \d g + \frac{\kappa}{4\pi} g' \d g \Big)\,. \nonumber
\end{align}
Here $\gamma$ is a path along the orbit. By parameterizing $\gamma$ by a coordinate $t$ on the orbit and using $\d = \d t \partial_t $ we find the two dimensional action
\begin{equation}\label{geomac}
I = \int \d t \d\vp \; \left( \cL_0 f' \dot f + \frac{c_{\rm eff}}{48\pi} \frac{f'' \dot f'}{f'{}^2} \right) + \int \d t \d\vp \; \left(\cP_0 f' \dot{\tilde g} + \frac{k}{4\pi} \tilde g' \dot{ \tilde g}\right) \,.
\end{equation}
Here dots denote $t$-derivatives and we have used the redefinition \eqref{redef} to remove the term proportional to $k$. The above action is expressed in terms of $\tilde g \circ f \circ \vp$. By redefining $f\circ \vp$ as $\tilde \vp $ and switching variables as $\tilde g(\tilde \vp) = \Phi(\vp)$, we may equivalently write the action above as:
\begin{equation}
I = \int \d t \d\vp \; \left( \cL_0 f' \dot f  + \frac{c_{\rm eff}}{48\pi} \frac{f'' \dot f'}{f'{}^2} + \frac{\kappa}{4\pi} \Phi' \dot{ \Phi} + \cP_0 \dot{\Phi} f' \right)\,.
\end{equation}

The Hamiltonian for this action can be found as the action of $L_0$ and $P_0$, i.e. we can add $- \int \d t (L_0 + P_0)$ to \eqref{geomac}, as in section 4 of \cite{Afshar:2019tvp} and in \cite{Henneaux:2019sjx} for the Virasoro group. This generalizes the warped Schwarzian action (4.5) of \cite{Afshar:2019tvp} to a generic orbit parameterized by $\mathcal{L}_0$ and $\mathcal{P}_0$. For further insights, we refer the reader to \cite{Barnich:2017jgw} for a discussion on invariant Hamiltonians for geometric actions. The key concept is that $L_0$ always belongs to the stabilizer subgroup of the orbit, thereby generating a gauge symmetry on the orbit. The generator of this gauge symmetry can then be included in the action as the Hamiltonian, preserving the gauge symmetry.

From \eqref{KK symp form}, we can make the field redefinition $\tilde{g} \rightarrow \tilde{g}-\frac{2\pi}{\kappa} \cP_0 f$ and get the following symplectic form:
\begin{equation}
	\omega = - \oint \d\vp \; \Big( (\cL_0-\frac{\pi}{\kappa} \cP_0^2) \d f \wedge \d f' + \frac{c_{\rm eff}}{48\pi} \d \log f'  \wedge (\d \log f')' + \frac{\kappa}{4\pi} \d g \wedge \d g' \Big).
\end{equation}
leading to the action
\begin{equation}
	I = \int \d t \d\vp \; \left( (\cL_0-\frac{\pi}{\kappa} \cP_0^2) f' \dot f  + \frac{c_{\rm eff}}{48\pi} \frac{f'' \dot f'}{f'{}^2} + \frac{\kappa}{4\pi} \Phi' \dot{ \Phi} \right)\,.
\end{equation}
Note that now, the field $\Phi$ is no longer periodic around the circle. So we need to make a last field redefinition $\Phi \rightarrow \Phi + \frac{2\pi}{\kappa} \cP_0 \varphi$ to get the final geometric action after dropping some boundary terms:
\begin{equation}
	I = \int \d t \d\vp \; \left( (\cL_0-\frac{\pi}{\kappa} \cP_0^2) f' \dot f  + \frac{c_{\rm eff}}{48\pi} \frac{f'' \dot f'}{f'{}^2} + \frac{\kappa}{4\pi} \Phi' \dot{ \Phi} +  \cP_0 \dot{\Phi} \right)\,.
\end{equation}
Adding the Hamiltonian
\begin{equation} \label{hamiltonian}
	H =   \mu L_0 + \nu P_0  \, ,
\end{equation}
where $\mu$ and $\nu$ are chemical potentials and $L_0$ and $P_0$ are the zero modes of $\cL(\varphi)$ and $\cP(\varphi)$ respectively:
\begin{equation}
	L_0 = \int \d\vp \left( (\cL_0-\frac{\pi}{\kappa} \cP_0^2) f'{}^2 + \frac{c_{\rm eff}}{48\pi} \frac{f''{}^2}{f'{}^2} + \frac{\kappa}{4\pi} \Phi'{}^2 + \frac{\pi}{\kappa} \cP_0^2 \right) \, , \quad P_0 = \int \d\vp \cP_0 \ ,
\end{equation} 
this leads to the action 
\begin{align}\label{geomac2}
    \begin{split}
	S &= I - \int \d t \, H \\
        &= \int \d t \d\vp \; \left( (\cL_0-\frac{\pi}{\kappa} \cP_0^2) f' \partial_- f  + \frac{c_{\rm eff}}{48\pi} \frac{f'' \partial_- f'}{f'{}^2} + \frac{\kappa}{4\pi} \Phi' \partial_- \Phi + \cP_0 \dot{\Phi} - \mu \frac{\pi}{\kappa} \cP_0^2 - \nu \cP_0  \right)\,.
    \end{split}
\end{align}
This action represents the geometric action on the coadjoint orbits of the warped Virasoro group with orbit representatives $(\cL_0, \cP_0)$. We expect it to match the reduction of the Chern-Simons theory with constant charges that we will compute in the next section.

\section{Chern-Simons reduction}
We start with a $SL(2,\mathbb{R}) \times U(1)$ Chern-Simons action:
\begin{equation}
	S_{CS} = \frac{k}{4 \pi} \int_\mathcal{M} \left< A \wedge \d A + \frac{2}{3} A \wedge A \wedge A \right> + \frac{\kappa}{4 \pi} \int_\mathcal{M}  \left< C \wedge \d C \right> \, ,
\end{equation}
written in Hamiltonian form:
\begin{equation}
	S_{CS} = \frac{k}{4 \pi} \int_\mathcal{M} \left< A_\varphi \dot{A}_r - A_r \dot{A}_\varphi + 2 A_t F_{r \varphi} \right> \d^3x + \frac{\kappa}{2 \pi} \int_\mathcal{M} ( C_\varphi \dot{C}_r + C_t \tilde{F}_{r \varphi }) \  \d^3x \, , \label{Ham form CS action}
\end{equation}
where $F = \d A + A \wedge A$ and $\tilde{F} = \d C$. We suppose that the topology of our manifold $\mathcal{M}$ is an annulus times $\mathbb{R}$, such that there are two boundaries at $r=r_i$ and $r=r_o$, respectively for the inner and the outer ones. On these boundaries, we impose the following boundary conditions for $\mu$ and $\nu$ constant:
\begin{align} \label{BC}
\begin{split}
	&\text{At} \ r=r_o : \left\{
		\begin{array}{l}
			A_\varphi = L_1 - \mathfrak{L} L_{-1} \\
			A_t = \mu A_\varphi
		\end{array} \right. \text{and} \ \left\{
		\begin{array}{l}
			C_\varphi = \frac{2 \pi}{\kappa} \mathcal{P} \\
			C_ t = \mu C_\varphi + \nu
		\end{array}\right. \\
	&\text{At} \ r=r_i : \left\{
		\begin{array}{l}
			A_\varphi = L_{-1} - \tilde{\mathfrak{L}} L_1 \\
			A_t = - \tilde{\mu} A_\varphi
		\end{array} \right. \text{and} \ \left\{
		\begin{array}{l}
			C_\varphi = \frac{2 \pi}{\kappa} \tilde{\mathcal{P}} \\
			C_ t = - \tilde{\mu} C_\varphi - \tilde{\nu}
		\end{array}\right. 
\end{split}
\end{align} 
with $\mathfrak{L} = \frac{2\pi}{k} (\mathcal{L} - \frac{\pi}{\kappa} \mathcal{P}^2)$ and $\tilde{\mathfrak{L}} = \frac{2\pi}{k} (\tilde{\mathcal{L}} - \frac{\pi}{\kappa} \tilde{\mathcal{P}}^2)$.
The equations of motion fix the time evolution of $\mathcal{L}$ and $\mathcal{K}$ for constant chemical potentials:
\begin{equation}
	\partial_t \mathcal{L} = \mu \mathcal{L}' \ , \ \partial_t \mathcal{P} = \mu \mathcal{P}' \, , \label{EOM C-S}
\end{equation}
where the prime denotes a derivative with respect to $\varphi$.

To have a well defined variational principle, we need to add boundary terms to the action:

\begin{eqnarray}
	I_{\Sigma_o} &=& -\frac{k}{4 \pi} \int_{\Sigma_o}  \mu\left<A_\varphi^2\right>\d^2x - \frac{\kappa}{4 \pi} \int_{\Sigma_o}  (\mu C_\varphi^2 + 2\nu C_\varphi) \ \d^2x \label{Outer bound act}\\
	&=&- \int_{\Sigma_o} (\mu \mathcal{L} + \nu \mathcal{P}) \ \d^2x \, , \nonumber \\
	I_{\Sigma_i} &=&- \frac{k}{4 \pi} \int_{\Sigma_i}  \tilde{\mu}\left<A_\varphi^2\right> \d^2x - \frac{\kappa}{4 \pi} \int_{\Sigma_i}  (\tilde{\mu} C_\varphi^2 + 2\tilde{\nu} C_\varphi) \ \d^2x \label{Inner bound act}\\
	&=&- \int_{\Sigma_i} (\tilde{\mu} \tilde{\mathcal{L}} + \tilde{\nu} \tilde{\mathcal{P}}) \ \d^2x \, . \nonumber
\end{eqnarray}
Our final action is then
\begin{equation}
	S[A,C] = S_{CS} + I_{\Sigma_i} + I_{\Sigma_o} \, . \label{Total action}
\end{equation}
Solving the constraints $F_{r\varphi }=0$ and $\tilde{F}_{r \varphi}=0$ and taking care of the holonomy, we get:
\begin{equation}
	\begin{array}{l l}
		A_r = G^{-1} \partial_r G \, , & A_\varphi = G^{-1} (\partial_\varphi + K(t)) G \, , \\
		C_r = \partial_r \lambda \, , & C_\varphi = \partial_\varphi \lambda + k_0(t) \, ,
	\end{array}
\end{equation}
where $G$ is a group element of $SL(2,\mathbb{R})$ and is periodic, as is $\lambda$. $K(t)$ is a function parametrizing the holonomy and taking value in the Lie algebra.

First, we will focus on the $U(1)$-part of the action \eqref{Total action}. The $U(1)$-part in \eqref{Ham form CS action} can be expressed as
\begin{equation}
	S[\Phi(t,\varphi),\Psi(t,\varphi),k_0(t)] = \frac{\kappa}{4 \pi} \int \left(\partial_\varphi \Phi \partial_t \Phi-\partial_\varphi \Psi \partial_t \Psi + 2 k_0 (\partial_t \Phi - \partial_t \Psi)\right) \d t \ \d\varphi \, ,
\end{equation}
where $\lambda(r=r_o,t,\varphi)=\Phi(t,\varphi)$ and $\lambda(r=r_i,t,\varphi)=\Psi(t,\varphi)$. From the boundary terms \eqref{Outer bound act} and \eqref{Inner bound act}, we have
\begin{equation}
	I_B = - \frac{\kappa}{4\pi} \int \left(\mu (\partial_\varphi \Phi)^2 + \tilde{\mu} (\partial_\varphi \Psi)^2 + (\mu + \tilde{\mu})k_0^2 + 2(\nu + \tilde{\nu}) k_0 \right) \d t \ \d\varphi \, .
\end{equation}
Using the boundary conditions \eqref{BC}, the functions $\Phi$ and $\Psi$ are constrained by
\begin{align}
	\begin{split}
		\partial_\varphi \Phi + k_0 &= \frac{2 \pi}{\kappa} \mathcal{P} \, ,\\
		\partial_\varphi \Psi + k_0 &= \frac{2 \pi}{\kappa} \tilde{\mathcal{P}} \, .
	\end{split}
\end{align}
So to summarize, we have
\begin{equation}
	S[\Phi(t,\varphi),\Psi(t,\varphi),k_0(t)] = S^{(2)} + S^{(1)} +S^{(0)} \, ,
\end{equation}
with
\begin{align}
\begin{split}
		S^{(2)} &= \frac{\kappa}{4 \pi} \int \d^2x \left(\partial_\varphi \Phi \partial_t \Phi -\mu (\partial_\varphi \Phi)^2 \right) \, , \\
		S^{(1)} &= \frac{\kappa}{4 \pi} \int \d^2x \left(-\partial_\varphi \Psi \partial_t \Psi - \tilde{\mu} (\partial_\varphi \Psi)^2 \right) \, , \\
		S^{(0)} &= \frac{\kappa}{2 \pi} \int \d^2x \left(k_0 (\partial_t \Phi - \partial_t \Psi) - \frac{\mu + \tilde{\mu}}{2}k_0^2 -(\nu + \tilde{\nu}) k_0 \right) \, .
\end{split}
\end{align}
This action is invariant under the $U(1)$ gauge transformation
\begin{equation}
    \Phi \to \Phi + \epsilon(t) \, , \qquad \Psi \to \Psi + \epsilon(t) \, , \qquad k_0 \to k_0 \, ,
\end{equation}
and under the $\frac{\widehat{LG} \times \widehat{LG}}{G}$ global symmetry with $G= U(1)$ \cite{Henneaux:2019sjx,ELITZUR1989108}
\begin{equation}
    \Phi \to \Phi + \alpha (\vp) \, , \qquad \Psi \to \Psi + \beta (\vp) \, ,
\end{equation}
where $\widehat{LG}$ is the loop group of a topological group $G$ \cite{DeKerf1997:Chap18}. The quotient by $G$ arises from the sharing of the same zero mode, related to the holonomy $k_0$, for $\Phi$ and $\Psi$. 

If we focus on one boundary only, we get:
\begin{equation}
	S_{\text{bdy}}^{\Sigma_o} = \frac{\kappa}{4 \pi} \int d^2x \left( \Phi' \partial_- \Phi + 2 k_0 \dot{\Phi} - \mu k_0^2 -2 \nu k_0 \right) \, ,
\end{equation}
where we have defined $\partial_- = \partial_t - \mu \partial_\varphi$ and dropped a total derivative at the time boundaries.
The remaining action has lost its gauge invariance and is left with only a $\widehat{LG}$ global symmetry.\\

Now, let us turn to the $SL(2,\mathbb{R})$ part of the action \eqref{Total action}. The computation was already done in \cite{Henneaux:2019sjx} for similar boundary conditions. The only difference is the presence of $\mu$ in the boundary conditions \eqref{BC} (set to $1$ in \cite{Henneaux:2019sjx}) and the replacement of $\cL$ by $\mathfrak L$.
In summary, we have at the outer boundary:
\begin{equation}
	S_{\text{bdy}}^{\Sigma_o} = \frac{k}{8\pi} \int_{\Sigma_o} \d t \d\varphi \left[\frac{\partial_-f' f''}{f^{\prime 2}} + \alpha \partial_-f f' \right] \, ,
\end{equation}
and
\begin{equation}
	\mathfrak{L} = \frac{\alpha}{4} f^{\prime 2} - \frac{1}{2} \{f;\varphi\} \, ,
\end{equation}
with $\alpha$ defined as
\begin{equation} \label{alpha holonomy}
	\begin{array}{c|c|c}
		\text{Holonomy} &  K(t) & \alpha \\
		\hline
		\text{Hyperbolic} & k_h(t) L_0& k_h^2  \\
		\text{Elliptic} & \frac{1}{2} k_e(t)(L_1 + L_{-1}) & -k_e^2  \\
		\text{Parabolic} &k_p(t) L_{-1}& 0 
	\end{array}
\end{equation}
A similar action holds for the inner boundary. 
The fields $\Phi$ and $f$ living on the outer boundary are coupled through the boundary conditions \eqref{BC}:
\begin{align} \label{final bound cond}
	\begin{aligned}
		\Phi' + k_0 &= \frac{2 \pi}{\kappa} \mathcal{P} \, , \\ 
	 	\frac{\alpha}{2} f^{\prime 2} - \{f;\varphi\}  &= \frac{4\pi}{k} (\mathcal{L} - \frac{\pi}{\kappa} \mathcal{P}^2) \, ,
	 	\end{aligned}
\end{align}
and follow the action
\begin{equation}
	S_{\text{bdy}}^{\Sigma_o} = \frac{k}{8\pi} \int_{\Sigma_o} \d^2x \left(\frac{\partial_-f' f''}{f^{\prime 2}} + \alpha \partial_-f f' \right) + \frac{\kappa}{4 \pi} \int \d^2x \left( \Phi' \partial_- \Phi + 2 k_0 \dot{\Phi} - \mu k_0^2 -2 \nu k_0 \right) \, . \label{bound action}
\end{equation}
We can now compare \eqref{bound action} with \eqref{geomac2} and observe that
\begin{equation} \label{L_0 P_0}
	\cP_0 = \frac{\kappa}{2\pi} k_0 \, , \qquad c_{\text{eff}} = 6k \, , \qquad \cL_0-\frac{\pi}{ \kappa} \cP_0^2 =  \frac{k}{8\pi} \alpha \, .
\end{equation}
The residual symmetries of \eqref{bound action} are a $U(1)$ gauge symmetry for the field $f$, in order to maintain the conjugacy class of the holonomy $\alpha$, as long as $\alpha \neq - n^2$ \cite{Henneaux:2019sjx} and $SL(2 , \mathbb R)^{(n)}$ when $\alpha = - n^2$, while the field $\Phi$ has a $\widehat{LG}$ global symmetry \cite{Henneaux:2019sjx}. All of this should be in concordance with the little group of the coadjoint orbits for constant representatives in the case $\kappa \neq 0$.

\section{Dimensional reduction and link with Warped Schwarzian theory}

The idea for this section is to perform the dimensional reduction of our action \eqref{bound action} as can be done for Liouville theory where one ends up with a Schwarzian theory \cite{Cotler:2018zff}.

The first step is to make a Wick rotation $t= i y$. Then we take the time circle to be very small, $\Delta y \rightarrow 0$, with $k' = k \Delta y$ and $\kappa' = \kappa \Delta y$. We then end up with
\begin{equation}
S =  \frac{\mu k'}{8 \pi} \int \d\varphi \left(\frac{ f^{\prime \prime 2}}{f^{\prime 2}} + \alpha f^{\prime 2}  \right) + \frac{\mu \kappa'}{4 \pi} \int \d\varphi \left( \Phi^{\prime 2}  + k_0^2 +2 \frac{\nu}{\mu} k_0 \right) \, .
\end{equation}
Now, up to some boundary terms, we can repackage this using the Schwarzian derivative \eqref{Scharzian derivative}:
\begin{eqnarray}
S &=& \frac{\mu k'}{8 \pi} \int \d\varphi \left(-2 \left\{f;\varphi\right\} + \alpha f^{\prime 2} \right) + \frac{\mu \kappa'}{4 \pi} \int \d\varphi \left( \Phi^{\prime 2}  + k_0^2 +2 \frac{\nu}{\mu} k_0 \right)  \\
&=& - \frac{\mu k'}{8 \pi} \int \d\varphi \left\{\tan\left( \frac{\sqrt{\alpha}}{2} f\right);\varphi\right\}  + \frac{\mu \kappa'}{4 \pi} \int \d\varphi \ \Phi^{\prime 2}  + \int \d\varphi \left(\mu \cL_0^{\prime} + \nu \cP_0^{\prime} \right) \, , \nonumber
\end{eqnarray}
where for the last step we used the equations \eqref{L_0 P_0}.
The last equation is the Warped Schwarzian action in \cite{Afshar:2019tvp} with additional source terms.

\section{One-loop partition function \texorpdfstring{\label{section warped 1-loop}}{}} 

We are now interested in computing the partition function for our action \eqref{bound action} in a similar manner as Section 5 of \cite{Cotler:2018zff}. 
The general idea behind this strategy was outlined in \cite{Maloney:2007ud, Giombi:2008vd, Maloney:2009ck}. The canonical ensemble partition function 
can be thought of an Euclidean functional integral weighted by the classical action with typical coupling constant $1/K$ (for AdS$_3$ gravity, this would be related to the Brown-Henneaux central charge as $c = 24K$). At large $K$ the dominant contribution to the path integral is given by the saddle point approximation, i.e. as a sum over classical saddles satisfying the appropriate boundary conditions of an exponential whose argument consist in a series expansion starting with the classical action evaluated on the saddle, followed by subleading terms representing quantum corrections to the effective action at n$^{th}$ order in perturbation theory (as it turns out, there are instances where the partition function happens to be one-loop exact). On general grounds, the partition function is expected to be expressible as a sum of characters of the corresponding symmetry algebra - Virasoro for AdS$_3$ gravity, and BMS$_3$ for flat space holography. This indeed turns out to be the case \cite{Giombi:2008vd, Cotler:2018zff, Merbis:2019wgk}. In the present situation the relevant characters are those of the warped conformal algebra taking the following form \cite{Apolo:2018eky} (see App.A for a brief review and notations):

\begin{align}
    \chi_{h,p} &= q^{h- \frac{c}{24}} y^p \prod_{n=1}^\infty \frac{1}{1-q^{2n}} \left( 1 - \delta_{\text{vac}} \, q \right) \quad  \,\,\,\, \text{for} \: p \in \mathbb R \, , \label{warped character p real}\\
    \chi_{h,p} &= q^{h- \frac{c}{24}} y^p \prod_{n=1}^\infty \frac{1}{(1-q^{n})^2} \left( 1 - \delta_{\text{vac}} \, q \right) \quad \text{for} \: p \in i \mathbb R \, .\label{warped character p imaginary}
\end{align}

We will therefore expand the partition function for our action \eqref{bound action} in orders of $k$ and $\kappa$ and limit ourselves to the 1-loop contribution. We refer the reader to \cite{Merbis:2019wgk} for a sketch of the proof of the one-loop exactness of the partition function for any geometric action defined from the Kirillov-Konstant symplectic form. We will investigate whether this allows us to recover the above characters.


From now on, we will assume that $\mu = 0$ and $\nu = 1$, as suggested in \cite{Azeyanagi:2018har}. This implies that the mass and angular momentum of the solutions with boundary conditions \eqref{Outer bound act} are given by
\begin{equation}
	M = \pi \cP \, , \qquad J = - 2 \pi \cL \, .
\end{equation}
The action \eqref{bound action} then reduces to 
\begin{equation}
	S_{\text{bdy}}^{\Sigma_o} = \frac{k}{8\pi} \int_{\Sigma_o} \d^2x \left(\frac{\dot f' f''}{f^{\prime 2}} + \alpha \dot f f' \right) + \frac{\kappa}{4 \pi} \int \d^2x \left( \Phi' \dot \Phi + 2 k_0 (\dot{\Phi}  -1 ) \right) \, . \label{bound action w/o chem pot}
\end{equation}

To compute the one-loop partition function, we first need the classical saddle point of the action \eqref{bound action w/o chem pot} and then evaluate the euclidean action on this saddle point. The equations of motion derived from \eqref{bound action w/o chem pot} for constant orbit representative are
\begin{align}
\begin{split}
	\frac{1}{f'} \partial_t \left(\frac{\alpha}{2} f^{\prime 2} - \{f;\varphi\}\right) &= 0 \, , \\
	\partial_t \Phi' &= 0 \, .
\end{split}
\end{align}
Using \eqref{final bound cond} and supposing $f' \neq 0$, we recover the equations of motion of the original Chern-Simons theory \eqref{EOM C-S} for our choice of chemical potentials. The solutions of interest to the saddle point approximation are those with constant $\mathfrak{L}$ and $\mathcal{P}$, where their values are given by the zero modes $\mathcal{L}_0$ and $\mathcal{P}_0$, implying from \eqref{final bound cond}
\begin{equation}
    \frac{\alpha}{2} \left( f^{\prime 2} -1 \right) = \{ f ; \vp \} \, , \qquad \Phi' = 0 \, .
\end{equation}
We get as saddle point the solution:
\begin{equation}
	f(\varphi,t) =  \varphi + c_1(t) \, , \qquad \Phi(\varphi,t) = c_2(t)  \, . 
\end{equation}
We make a Wick rotation $t=-i y$ and compute the euclidean action $iS_E = S_{\text{bdy}}^{\Sigma_o}$,
\begin{equation} \label{euclidean action real P0}
	S_E = - i \frac{k}{8\pi} \int_{\Sigma_o} \d^2x \left(  \frac{\dot{f'} f''}{f^{\prime 2}} +  \alpha \dot{f} f' \right) - \frac{\kappa}{4 \pi} \int \d^2x \left(i \Phi' \dot{\Phi} + 2i k_0 \dot{\Phi} - 2 k_0 \right) \, .
\end{equation}

For this part, we will separate the discussion into two cases: the first one for a real orbit representative $\cP_0$ and the second one for a purely imaginary $\cP_0$. Let us begin with a real $U(1)$ holonomy, which corresponds to warped black hole solutions, and derive the Hamiltonian as the real part of the action
\begin{equation} \label{Hamiltonian real P_0}
	H   = \kappa k_0 = 2 \pi \cP_0 \, ,
\end{equation}
which could already be seen from \eqref{hamiltonian} with $\mu =0$ and $\nu =1$.
It corresponds to an orbit representative $\cP_0$ that is also real.
For a given value of the holonomy, the Hamiltonian is always bounded from below. It is interesting to note that in the case of $\mu =0$ and $\nu=1$, the path integral is well-defined for any value of $\alpha$, as long as $k_0$ is real, contrary to the Virasoro case where only values of $\alpha$ greater or equal to $-1$ were allowed. 

Then, we periodically identify the euclidean time $y$. We get as periodicity conditions on the fields:
\begin{equation} \label{periodicity}
	\begin{aligned}
		f(\varphi+\Omega\beta,y+\beta) = f(\varphi,y) \qquad &,  \qquad \Phi(\varphi+\Omega\beta,y+\beta)= \Phi(\varphi,y) \, , \\
		f(\varphi+2\pi,y) = f(\varphi,y)+2 \pi  \qquad &,  \qquad \Phi(\varphi+2\pi,y) = \Phi(\varphi,y) \, .
	\end{aligned}
\end{equation} 
In terms of the fields $c_1(y)$ and $c_2(y)$, this gives
\begin{equation}
	c_1(y+\beta) = c_1(y)- \Omega \beta  \qquad ,  \qquad c_2(y+\beta) = c_2(y) \, .
\end{equation}
So the solution is
\begin{equation} \label{saddle point}
	f_0 = \varphi - \Omega y \, , \quad \Phi_0 = 0 \, ,
\end{equation}
and the euclidean action evaluated on this saddle point \eqref{saddle point} is
\begin{equation}
	S_E^{(0)} = \beta \left( \kappa k_0 + i k \Omega \mathfrak{L}_0   \right) \, .
\end{equation}
From there, we can already state that the 0-loop contribution will give the factors
\begin{equation}
    e^{-S_E^{(0)}} = q^{-\frac{c \alpha}{24}} \,  y^{i \kappa k_0}
\end{equation}
where we defined
\begin{equation}
    q = e^{2\pi i \tau} \, , \qquad y = e^{2\pi i z} \, ,
\end{equation}
with $\beta \Omega = 2 \pi \tau$ and $\beta = 2 \pi z$.
Furthermore, we are imposing that $\beta$ and $\Omega$ are determined in terms of the holonomies $\alpha$ and $k_0$ \cite{Azeyanagi:2018har}
\begin{equation}
	\text{exp}\left[ \frac{i\beta \Omega}{2\pi} \int A_\varphi \d\varphi \right] = - \mathbb{1} \,  , \quad  \text{exp}\left[\frac{i\beta}{2 \pi} \left(\int C_t \d \vp + \Omega \int C_\varphi \d\varphi \right)\right] = e^{2\pi i \gamma} \, ,
\end{equation}
leading to the following relations between $\beta$, $\Omega$ and the zero-modes:
\begin{equation}
	\beta = 2\pi \left( \gamma - \frac{\pi \cP_0}{\kappa \sqrt{\mathfrak L}_0} \right) \qquad , \qquad \Omega = \frac{1}{2\gamma \sqrt{\mathfrak L}_0 - \frac{2\pi \cP_0}{\kappa }} \qquad , \qquad \beta \Omega = \frac{\pi }{\sqrt{\mathfrak L}_0} \, . \label{beta omega}
\end{equation}
The $\gamma$ is an arbitrary constant and represents a kind of deformation at the level of the holonomy \cite{Azeyanagi:2018har}. If it is an integer, one recovers a similar result to BTZ with an holonomy living in the center of the gauge group. However, it is expected that it is not necessarily the case for warped spacetimes. For instance, for the warped black hole in canonical ensemble \cite{Azeyanagi:2018har, Anninos:2008fx},
\begin{equation}
    \gamma = \frac{2 \nu}{\nu^2+3} \, .
\end{equation}

Now we can expand the fields $f(\varphi,y)$ and $\Phi(\varphi,y)$ around the saddle points as
\begin{equation}
	\begin{aligned}
		f(\varphi,y)&= f_0 + \sum_{m,n} \frac{f_{m,n}}{(2\pi)^2} e^{\frac{2\pi i m y}{\beta}} e^{in f_0} \, , \\
		\Phi(\varphi,y) &= \sum_{m,n} \frac{\Phi_{m,n}}{(2\pi)^2} e^{\frac{2\pi i m y}{\beta}} e^{in f_0} \, .
	\end{aligned}
\end{equation}
One can check that these series expansions satisfy the periodicity conditions \eqref{periodicity}. Plugging this into the euclidean action, we obtain
\begin{equation}
	S_E = S_E^{(0)} - \frac{i}{16\pi^3} \sum_{m=-\infty}^\infty \sum_{n \neq 0} n (m-\tau n) \left( \frac{k}{2} |f_{m,n}|^2 (n^2+\alpha)  + \kappa  |\Phi_{m,n}|^2 \right) + ... \, ,
\end{equation}
where $f_{m,n}^*= f_{-m,-n}$, $\Phi_{m,n}^*= \Phi_{-m,-n}$ and the dots contains terms of the third order and more on the fields. One does not need to take the vacuum into account in this case since in holographic WCFTs it possesses an imaginary value for $\cP_0$\cite{Azeyanagi:2018har, Apolo:2018eky}.

We can now compute the one-loop partition function by integrating out the fields $f_{m,n}$ and $\Phi_{m,n}$:
\begin{equation} \label{1 loop partition function determinant}
	Z_{\text{1-loop}} = \int \cD f \cD \Phi \, e^{-S_E} = \cN e^{-S_E^{(0)}} \prod_{m,n} n^{-1} (m-\tau n)^{-1} (n^2+\alpha)^{-1/2} \, ,
\end{equation}
where $\cN$ is a normalization constant independent of $\beta$ and $\Omega$.
To perform the product for the $m$'s, we take the derivative of the logarithm of the partition function with respect to $\tau$:
\begin{eqnarray*}
	\partial_\tau \log Z_{\text{1-loop}} &=& - 2 \pi i  k \mathfrak{L}_0 + \sum_{n \neq 0} \sum_{m=-\infty}^\infty \frac{n}{m-\tau n} \\
	&=& - 2 \pi i  k \mathfrak{L}_0 - \sum_{n \neq 0} \left( \frac{1}{\theta} + 2 n \sum_{m=1}^\infty \frac{\tau n}{\tau^2 n^2 -m^2} \right) \\
	&=&  - 2 \pi i  k \mathfrak{L}_0 - \pi \sum_{n \neq 0} n \cot(n\pi\tau) \\
	&=& - 2 \pi i  k \mathfrak{L}_0 - 2 \pi \sum_{n =1 }^\infty n \cot(n\pi \tau)
\end{eqnarray*}
for $n\tau \notin \mathbb{Z}$. This sum diverges but there are several ways to deal with it. One is to use the zeta function regularization to write 
\begin{equation} \label{zeta regularization}
	\sum_{n =1 }^\infty n \cot(n\pi \tau) = \sum_{n =1 }^\infty n (\cot(n\pi \tau)+i) - i \sum_{n =1 }^\infty n \, .
\end{equation}
Now, the first sum on the right hand side converges for $\text{Im}(\tau) > 0$. This requirement is not a problem for negative value of $\alpha$, such as the vacuum. However for a positive value of the holonomy, one needs to perform an analytic continuation $\tau \to \tau + i \varepsilon$ as done in \cite{Barnich:2015mui}.
After exponentiation, the partition function reads (up to a normalization constant)
\begin{equation}
	Z_{\text{1-loop}} = e^{-S_E^{(0)}} q^{-\frac{1}{24}} \prod_{n=1}^\infty \frac{1}{(1-q^n)^2} = q^{-k \mathfrak{L}_0 - \frac{1}{24}} y^{i\kappa k_0}\prod_{n=1}^\infty \frac{1}{(1-q^n)^2} \, . \label{1-loop part func}
\end{equation}
The 1-loop partition function for a given saddle reproduces the contribution to the full WCFT partition function \eqref{warped character p imaginary}, originally computed in \cite{Apolo:2018eky} for the precise value $\kappa = -1$,


\begin{equation}
	\chi_{h,p}(\tau ,z) = q^{h-c/24} y^p \prod_{n=1}^\infty \frac{1}{(1-q^n)^2} \, ,
\end{equation}
for $p\in i \mathbb{R}$ which is our case here (see \eqref{1-loop part func}). From there we read that 
\begin{equation} \label{weigh charge real P0}
	h = - 2 \pi \left(\cL_0-\frac{\pi}{ \kappa} \cP_0^2 \right) + \frac{c-1}{24} \qquad , \qquad p = 2\pi i \cP_0 \, .
\end{equation}
It thus seems that starting from a real $\cP_0$, solutions representing black holes, we recover the partition function of a primary field with imaginary charge $p$, like the vacuum, and its descendants.



Such a result may seem unusual, but a similar phenomenon was discussed in \cite{Castro:2017mfj}, where the authors computed the one-loop partition function in AdS$_3$ with CSS boundary conditions \cite{Compere:2013bya} using a quasinormal mode method \cite{Denef:2009kn}. Their primary focus was to describe the BTZ black hole, a unitary representation, but they unexpectedly obtained the vacuum character, also a unitary representation in this context, which they associated with thermal AdS. They argued that, akin to CFTs, an appropriate modular transformation relates thermal AdS to BTZ \cite{Detournay:2012pc, Song:2016gtd}, rendering them indistinguishable because they both derive from Euclidean AdS \cite{Banados:1992gq, Carlip:1994gc, Maldacena:1998bw}. The only distinguishing factor is their contractible or non-contractible thermal circle, which explains why they obtained the vacuum determinant while computing the BTZ determinant.
Another possible interpretation comes from the analytic continuation procedure. Indeed, the Wick rotation $t = -i y$ has the effect of sending $\mathcal{P}_0$ to $i \mathcal{P}_0$, hence swapping the reality property of $\mathcal{P}_0$ and $p$. \\


We are now interested in states with purely imaginary values of $\cP_0$, such as the vacuum. Indeed, in that case, from \cite{Azeyanagi:2018har}, $\alpha = -  (1+2j)^2$ where $j$ is an integer and $k_0 = i \gamma$. We choose the branch $j=0$ as in \cite{Azeyanagi:2018har} to make the link with the WCFT vacuum values in \cite{Detournay:2012pc}. When the orbit representative $\cP_0$ is purely imaginary, $\mathfrak L_0$ remains real and the euclidean action \eqref{euclidean action real P0} is instead
\begin{equation} \label{euclidean action imaginary P0}
	S_E = -\frac{ik}{2\pi} \int_{\Sigma_o} \d^2x \left( \frac{\dot{f'} f''}{4f^{\prime 2}} +  \mathfrak{L}_0 \dot{f} f' \right) -  \int \d^2x \left(\frac{i\kappa}{8 \pi} \Phi' \dot{\Phi} + i \cP_0 \dot{\Phi} - \cP_0 \right) \, .
\end{equation}
The real part of the action is then
\begin{equation} \label{real part action imaginary P0}
    \text{Re}[S_E] = - i \int \d \vp \cP_0 \dot{\Phi} \, .
\end{equation}
Since we are now dealing with a non-unitary representation, \eqref{real part action imaginary P0} can no longer be interpreted as the Hamiltonian \eqref{hamiltonian}. For the field periodicity \eqref{periodicity}, a saddle point of the action \eqref{euclidean action imaginary P0} is still given by \eqref{saddle point}. One can show that any perturbations around this saddle point keep the real part of the action \eqref{real part action imaginary P0} bounded, such that the path integral derivation can also be performed. Following the previous computations, the one-loop partition function is simply
\begin{equation}
	Z_{\text{1-loop}} = q^{-k \mathfrak{L}_0 - \frac{1}{24}} y^{-\kappa k_0}\prod_{n=1}^\infty \frac{1}{(1-q^n)^2} \, . 
\end{equation}
which is \eqref{1-loop part func} with $k_0 \rightarrow i k_0$. 
This reproduces the character 
for an imaginary one \eqref{warped character p imaginary}. 

For the vacuum, the computation of the $SL(2,\mathbb{R})$ part of the partition function needs to be improved because $\alpha = -1$ in \eqref{1 loop partition function determinant}, making start the product at $n=2$. It implies a slightly modified zeta regularization \eqref{zeta regularization}
\begin{equation} \label{zeta regularization vacuum}
    \sum_{n=2}^\infty n \rightarrow \zeta (-1) - 1 = - \frac{13}{12} \, .
\end{equation}
The one-loop partition function for the vacuum, $\alpha = -1$ and $k_0 = i \gamma$, is thus
\begin{equation}
	Z_{\text{1-loop}}^{\text{vac}} = q^{- \frac{c+13}{24}} y^{-\kappa \gamma} \prod_{n=1}^\infty \frac{1}{(1-q^{n})^2} (1- \delta_{\text{vac}} q) \, .
\end{equation}
The factor $13$ raised from the zeta regularization because we wanted to "regularize first, integrate later". In \cite{Barnich:2015mui, Merbis:2019wgk}, a different approach, "integrate first, regularize later", was performed which prevents the apparition of the regularization factor.


We surprisingly observe that this computation again yielded the character for $p$ imaginary, as was obtained for real $\mathcal{P}_0$. How is it possible to obtain the characters for $p$ real?
In order to address this apparent puzzle, we momentarily pause on some particularities of WCFT partition functions. We have so far implicitly assumed that the bulk and boundary theories are defined by the following partition function
\begin{equation}
Z^{\text{CE}}(\beta,\theta) = \mbox{Tr} \; e^{-\beta P_0  + i \theta L_0}
\end{equation}
which is said to be in \emph{canonical} ensemble, and where $L_0$ and $P_0$ are the zero modes of the algebra \eqref{WCFT algebra}. It appears however more natural in some situations \cite{Detournay:2012pc} to define the following generators and partition function:
\begin{equation}\label{quadratic ensemble relation to canonical}
L_n^Q=L_n-\frac{2}{\mathsf{k}}P_0P_n + \frac{1}{\mathsf{k}}P_0^2\delta_n~,\quad P_n^Q=-\frac{2}{\mathsf{k}}P_0P_n +\frac{1}{\mathsf{k}}P_0^2\delta_n~.
\end{equation} 
and
\begin{equation} \label{QPF}
  Z^{\text{QE}}(\beta_R,\beta_L) = \mbox{Tr} \; e^{-\beta_R P_0^Q  - \beta_L L_0^Q}~.
\end{equation} 
The latter defines the so-called \emph{quadratic} ensemble partition function. Importantly, we have in particular that $P_0^Q = -P_0^2/k$, hence states with opposite values of $P_0$ are identified.
We will refrain from reconsidering our previous analysis in quadratic ensemble, which would entail reformulating the appropriate bulk boundary conditions \cite{Aggarwal:2020igb} in Chern-Simons/Lower-Spin Gravity language, then going through the steps of the Hamiltonian reduction, and identifying the corresponding geometric action. We will leave this for further work. However, it is natural to expect that the periodic boundary conditions on the quadratic ensemble $U(1)$ field, when expressed in terms of the canonical $U(1)$ field above, will lead to either periodic, \emph{or anti-periodic} boundary conditions. Let us inspect the consequences of this observation, by considering the following (anti-)periodicity condition: 


\begin{equation} \label{antiperiod cond}
	\Phi(\varphi+\Omega\beta,y+\beta)= - \Phi(\varphi,y) \, .
\end{equation}


The new saddle point satisfying this antiperiodic condition \eqref{antiperiod cond} can be written in the form
\begin{equation}
    \Phi_0 = \sin \left( \frac{\pi y}{\beta} \right)  \, ,
\end{equation}
and the expansion of the field around his saddle point is
\begin{equation} \label{expansion saddle imag P0}
	\Phi(\varphi,y) = \Phi_0 + \sum_{m,n} \frac{\Phi_{m,n}}{(2\pi)^2} e^{\frac{2\pi i m y}{\beta}+\frac{i \pi y}{\beta}} e^{in f_0} \, .
\end{equation}
The real part of the action \eqref{real part action imaginary P0} is also bounded for the perturbations \eqref{expansion saddle imag P0} and we can perform the path integral derivation.
Focusing on the $U(1)$ part of the action, the euclidean action around the saddle point is
\begin{equation}
	S_E^{U(1)} = \kappa \beta k_0 + \frac{i}{2} \sum_{m=-\infty}^\infty \sum_{n \neq 0}  \kappa  \frac{|\Phi_{m,n}|^2}{(2\pi)^3} n \left(m-\tau n - \frac{1}{2} \right) + ... \, .
\end{equation}
The one-loop $U(1)$ partition function becomes
\begin{equation}
	Z_{\text{1-loop}}^{U(1)} = \cN y^{i\kappa k_0} \prod_{m,n} n^{-1/2} \left(m-\tau n- \frac{1}{2} \right)^{-1/2} \, ,
\end{equation}
for $n \neq 0$. Using the same trick as in Section \ref{section warped 1-loop}, we derive the logarithm of the partition function according to $\tau$ to perform the sum over m, then we integrate and exponentiate to get the one-loop partition function. First we perform the logarithm and rewrite the sum as
\begin{eqnarray*}
	\log Z_{\text{1-loop}}^{U(1)} &=& -\frac{1}{2} \sum_{n \neq 0} \sum_{m=-\infty}^\infty \log\left(m-\tau n-\frac{1}{2}\right) \\
	&=& -\frac{1}{2} \sum_{n=1}^\infty \sum_{m=-\infty}^\infty \left[ \log\left(m-\tau n-\frac{1}{2}\right) + \log\left(m+\tau n-\frac{1}{2}\right] \right)\\
	&=& -\frac{1}{2} \sum_{n=1}^\infty \sum_{m=-\infty}^\infty \log\left(\frac{1}{4}-m+m^2-n^2\tau^2\right) \, .
\end{eqnarray*} 
Now, deriving according to $\tau$:
\begin{eqnarray*}
	\partial_\tau \log Z_{\text{1-loop}}^{U(1)} &=& \sum_{n \neq 0} \sum_{m=-\infty}^\infty \frac{n^2 \tau}{\frac{1}{4}-m+m^2-n^2\tau^2} \\
	&=& \sum_{n \neq 0} \left( \sum_{m=-\infty}^{-1} \frac{n^2 \tau}{\frac{1}{4}-m+m^2-n^2\tau^2} + \frac{n^2 \tau}{\frac{1}{4}-n^2\tau^2} + \sum_{m=1}^{\infty} \frac{n^2 \tau}{\frac{1}{4}-m+m^2-n^2\tau^2} \right) \\
	&=& \sum_{n \neq 0} \left( \sum_{m=1}^{\infty} \frac{n^2 \tau}{\frac{1}{4}+m+m^2-n^2\tau^2} + \frac{n^2 \tau}{\frac{1}{4}-n^2\tau^2} + \sum_{m=1}^{\infty} \frac{n^2 \tau}{\frac{1}{4}-m+m^2-n^2\tau^2} \right) \\
	&=& \sum_{n \neq 0} \left( \frac{\pi}{2} \tan(n\pi \tau) + \frac{n^2 \tau}{n^2\tau^2 - \frac{1}{4}} + \frac{n^2 \tau}{\frac{1}{4}-n^2\tau^2} + \frac{\pi}{2} \tan(n\pi \tau) \right) \\
	&=& \sum_{n \neq 0} \pi \tan(n\pi \tau) \, .
\end{eqnarray*} 
Again this sum diverges but can be regularized using the zeta function regularization. After integration and exponentiation (up to a normalization constant), the $U(1)$ part of the one-loop partition function is
\begin{equation}
	Z_{\text{1-loop}}^{U(1)} = \cN' y^{i\kappa k_0} \prod_{n=1}^\infty \frac{1}{1+q^n} \, ,
\end{equation}
which is related to the $U(1)$ part of the Virasoro-Ka\v{c}-Moody character in (3.11) of \cite{Apolo:2018eky}. Combining this with the $SL(2,\mathbb{R})$ part that remains the same, we end up with
\begin{equation}
	Z_{\text{1-loop}} = q^{-k \mathfrak{L}_0 - \frac{1}{24}} y^{i\kappa k_0} \prod_{n=1}^\infty \frac{1}{(1-q^{2n})} \, ,
\end{equation}

For the vacuum, $\alpha=-1$ and $k_0 = i \gamma$. Once again, the computation of the $SL(2,\mathbb{R})$ part of the partition function requires improvement. Paying attention to the zeta regularization, as in \eqref{zeta regularization vacuum}, the one-loop partition for the vacuum with anti-periodic boundary conditions on the $U(1)$ field $\Phi$ \eqref{antiperiod cond} is 
\begin{equation}
	Z_{\text{1-loop}}^{\text{vac}} = q^{- \frac{c+13}{24}} y^{-\kappa \gamma} \prod_{n=1}^\infty \frac{1}{(1-q^{2n})} (1- \delta_{\text{vac}} q) \, .
\end{equation}
It is completely equivalent to the Warped Virasoro character \eqref{warped character p real} 
with weight and real charge (since $P_0$ is purely imaginary) \eqref{weigh charge real P0}.\\


We conclude from this section that using the periodic boundary conditions \eqref{periodicity} and computing the one-loop partition function of a unitary representation ($\mathcal{P}_0$ real), like a black hole, we get the warped character for an imaginary 
charge $p$. 
On the other hand, the one-loop determinant for fluctuations around states with imaginary charge (which includes the vacuum for holographic WCFTs) yields the warped character for a real charge $p$ only when anti-periodic boundary conditions (\ref{antiperiod cond}) on the canonical $U(1)$ field are used. Actually, periodic/anti-periodic boundary conditions on states with arbitrary (real or purely imaginary $\mathcal{P}_0$) result in a character for imaginary/real $p$. Though we do not have a clear interpretation for this result, we believe it would be interesting to reconsider our analysis for WCFTs in quadratic ensemble, or to use alternative methods (such as the Heat Kernel method \cite{Giombi:2008vd} or the quasinormal modes approach \cite{Castro:2017mfj}) to compute the one-loop partition function and confirm our findings.




\section*{Acknowledgements}
The authors are very grateful to Wout Merbis for collaboration at an early stage of this work and for many valuable contributions and discussions.
They thank Luis Apolo and Max Riegler for useful discussions and exchanges on the topics covered in this work. SD is a Senior Research Associate of the Fonds de la Recherche Scientifique F.R.S.-FNRS (Belgium). SD and QV were supported in part by IISN – Belgium (convention 4.4503.15) and benefited from the support of the Solvay Family. SD acknowledge support of the Fonds de la Recherche Scientifique F.R.S.-FNRS (Belgium) through the PDR/OL C62/5 project “Black hole horizons: away from conformality” (2022-2025).

\appendix

\section{Warped Virasoro characters}

Modular invariance, or more precisely covariance in our case, can be used to derive the Warped Cardy formula. As stated in \cite{Apolo:2018eky}, it can also be used to derive the warped characters as we now review.


Let us start with the WCFT partition function
\begin{equation}
    Z (\tau , z) = \tr \left( q^{L_0-\frac{c}{24}} y^{P_0} \right) \, ,
\end{equation}
with
\begin{equation}
    q = e^{2\pi i \tau} \, , \qquad y = e^{2\pi i z} \, .
\end{equation}
Written is such a way, we can define two transformations generating an $SL(2 ,\mathbb Z)$ group \cite{Detournay:2012pc}
\begin{align}
    S &: \qquad \tau \to - \frac{1}{\tau} \, , \quad z \to \frac{z}{\tau} \, , \\
    T &: \qquad \tau \to \tau +1 \, .
\end{align}
The S-transformations correspond to modular transformations while the T-transformations represent the addition of the angular circle to the thermal circle. For instance, the S-transformations act on the partition function as
\begin{equation}
    Z\left(-\frac{1}{\tau} , \frac{z}{\tau}\right) = e^{-\frac{i \pi}{2} \frac{z^2}{\tau}} Z(\tau, z) \, . 
\end{equation}
\\

As with CFTs, the Hilbert space decomposes into a sum over highest-weight representations of primary states $|p , h \rangle$, defined as
\begin{align} \label{primary state}
\begin{split}
    P_n \left|p,h \right> = 0  \quad n > 0 \, , &\qquad L_n \left|p,h \right> = 0 \quad n > 0 \, , \\
    P_0 \left|p,h \right> = p \left|p,h \right> \, , &\qquad L_0 \left|p,h \right> = h \left|p,h \right> \, .
\end{split}
\end{align}
It implies a decomposition of the partition function into warped Virasoro characters
\begin{equation}
    Z(\tau , z) = \sum_{h,p} d_{h,p} \, \chi_{h,p} (\tau, z) \, ,
\end{equation}
where the sum runs only over primary states and $d_{h,p}$ takes into account their degeneracy. We will follow the path done in \cite{Apolo:2018eky} to compute the warped character of a primary state $|p , h \rangle$ \eqref{primary state}.

First, we use the Sugawara basis
\begin{equation} \label{sugawara basis}
    L_n^{(s)} = L_n - \frac{1}{k} \sum_{m=-\infty}^{\infty} \, : P_m P_{n-m} \, : \quad ,
\end{equation}
to compute the contribution from the Virasoro descendants to the partition function. The advantage of this basis is that it commutes with every $P_n$, allowing us to factorize the norm of mixed states admitting both Virasoro and $U(1)$ descendants. As a result, the warped Virasoro character is simply the product of both contributions. The Virasoro descendants are thus written as
\begin{equation} \label{sugawara descendants}
    \prod_{k=1}^\infty \left( L_{-k}^{(s)} \right)^{n_k} |p, h \rangle \, ,
\end{equation}
with $n_k$ integer characterising the descendant. As the commutator between $L_0$ and the Sugawara basis is
\begin{equation}
    [L_0 , L_n^{(s)}] = - n L_n^{(s)} \, ,
\end{equation}
the descendants \eqref{sugawara descendants} are proportional to
\begin{equation}
     \prod_{k=1}^\infty \left( L_{-k}^{(s)} \right)^{n_k} |p, h \rangle \propto |p , h + N \rangle \, .
\end{equation}
where $N = \sum_{k} n_k , k$. The degeneracy of the states $|p , h + N \rangle$ corresponds to the number of partitions of the positive integer $N$, often denoted as $p(N)$. For instance, $p(4) = 5$ because there are five different ways to partition 4: $1+1+1+1$, $1+1+2$, $2+2$, $3+1$, and $4$. Under the condition that the conformal weight $h$ of the primary state and the central charge $c$ satisfy
\begin{equation} \label{negative k requirement}
    c > 1 \, , \qquad h \geq \frac{p^2}{k} \ , \qquad \text{with} \: k < 0 \, ,
\end{equation}
the contribution of the Virasoro descendants is
\begin{equation}
    \sum_{N=1}^\infty p(N) \, q^N = \prod_{n=1}^\infty \sum_{k=0}^\infty q^{nk} =   \prod_{n=1}^\infty \frac{1}{1-q^n} \, ,
\end{equation}
like for CFTs.
When the primary state is the vacuum $|p,0\rangle$, the charge $L_{-1}$ acts on the vacuum as
\begin{equation}
    L_{-1}^{(s)} |p,0\rangle = 0 \, ,
\end{equation}
implying a state of vanishing norm that we need to exclude from the partition function by starting the product at $n=2$. This can be summarized as
\begin{equation}
    \prod_{n=1}^\infty  \frac{1}{1-q^n} \left( 1 - \delta_{\text{vac}} \, q \right) \, ,
\end{equation}
where $\delta_{\text{vac}} = 1$ for the vacuum and 0 otherwise.

For the $U(1)$ descendants, we need to separate the discussion into two cases. If the charge $p$ is real, the $P_n$ are Hermitian and the $U(1)$ descendants have two distinct norms, after an appropriate normalization: 
\begin{equation} \label{u(1) descendants}
    \left| \prod_{k=1}^\infty P_{-k}^{n_k} |p, h \rangle \right|^2 = \left\{
    \begin{array}{l}
         +1 \quad \text{if} \: \sum_k n_k \, \text{is even} \, ,  \\
         -1 \quad \text{if} \: \sum_k n_k \, \text{is odd} \, .
    \end{array}
    \right.
\end{equation}
The action of the $P_{-n}$s does not modify the charge $p$ but changes the weight since
\begin{equation}
    [L_0 , P_n] = - n P \, ,
\end{equation}
implying that the $U(1)$ descendants are proportional to the states of weight $h + N$
\begin{equation}
    \prod_{k=1}^\infty P_{-k}^{n_k} |p, h \rangle \propto |p , h + N \rangle \, .
\end{equation}
Unlike the Virasoro descendants, the degeneracy of the state $|p , h + N \rangle$ is not simply the partition function $p(N)$ because, due to the presence of negative norms in \eqref{u(1) descendants}, we also need to take into account the parity of the number of integers used to build the integer $N$. We denote this function as $f(N)$. For example, $f(4) = 1$ because $1+1+1+1$, $2+2$, and $3+1$ contain an even number of integers, contributing $+3$, but $1+1+2$ and $4$ use an odd number of integers, contributing $-2$. The total sum is $3 + (-2) = 1$. The contribution of the $U(1)$ descendants to the partition function is then\footnote{A rigorous proof of the computation of the generating function of $f(N)$ can be done by adapting the proof done for the integer partition function $p(N)$ in Section 3.3 of \cite{Guichard}.}
\begin{equation} \label{f(N) generating function}
    \sum_{N=1}^\infty f(N) \, q^N = \prod_{n=1}^\infty \sum_{k=0}^\infty (-1)^k q^{nk} =   \prod_{n=1}^\infty \frac{1}{1+q^n} \, .
\end{equation}
Since no $P_{-n}$ annihilates the vacuum, this result does not change when the primary state is the vacuum state.

The second possibility for the primary state is to possess a imaginary charge $p$. In this situation, the $P_n$s are antihermitians but their corresponding $U(1)$ descendants always have a positive norm. Their contribution to the partition will then take the same form as for the Virasoro descendants, without the discussion about the vacuum,
\begin{equation}
    \sum_{N=1}^\infty p(N) \, q^N =   \prod_{n=1}^\infty \frac{1}{1-q^n} \, ,
\end{equation}

In summary, the warped Virasoro characters are
\begin{align}
    \chi_{h,p} &= q^{h- \frac{c}{24}} y^p \prod_{n=1}^\infty \frac{1}{1-q^{2n}} \left( 1 - \delta_{\text{vac}} \, q \right) \quad  \,\,\,\, \text{for} \: p \in \mathbb R \, , 
    \\
    \chi_{h,p} &= q^{h- \frac{c}{24}} y^p \prod_{n=1}^\infty \frac{1}{(1-q^{n})^2} \left( 1 - \delta_{\text{vac}} \, q \right) \quad \text{for} \: p \in i \mathbb R \, .
\end{align}

\bibliography{biblio}

\providecommand{\href}[2]{#2}\begingroup\raggedright\begin{thebibliography}{10}

\bibitem{Hofman:2011zj}
D.~M. Hofman and A.~Strominger, ``{Chiral Scale and Conformal Invariance in 2D
  Quantum Field Theory},'' {\em Phys.Rev.Lett.} {\bf 107} (2011) 161601,
\href{http://www.arXiv.org/abs/1107.2917}{{\tt 1107.2917}}.

\bibitem{Detournay:2012pc}
S.~Detournay, T.~Hartman, and D.~M. Hofman, ``{Warped Conformal Field
  Theory},'' {\em Phys.Rev.} {\bf D86} (2012) 124018,
\href{http://www.arXiv.org/abs/1210.0539}{{\tt 1210.0539}}.

\bibitem{Castro:2015uaa}
A.~Castro, D.~M. Hofman, and G.~S\'arosi, ``{Warped Weyl fermion partition
  functions},'' {\em JHEP} {\bf 11} (2015) 129,
  \href{http://www.arXiv.org/abs/1508.06302}{{\tt 1508.06302}}.

\bibitem{Song:2017czq}
W.~Song and J.~Xu, ``{Correlation Functions of Warped CFT},'' {\em JHEP} {\bf
  04} (2018) 067, \href{http://www.arXiv.org/abs/1706.07621}{{\tt 1706.07621}}.

\bibitem{Apolo:2018eky}
L.~Apolo and W.~Song, ``{Bootstrapping holographic warped CFTs or: how I
  learned to stop worrying and tolerate negative norms},'' {\em JHEP} {\bf 07}
  (2018) 112, \href{http://www.arXiv.org/abs/1804.10525}{{\tt 1804.10525}}.

\bibitem{Castro:2015csg}
A.~Castro, D.~M. Hofman, and N.~Iqbal, ``{Entanglement Entropy in Warped
  Conformal Field Theories},'' {\em JHEP} {\bf 02} (2016) 033,
  \href{http://www.arXiv.org/abs/1511.00707}{{\tt 1511.00707}}.

\bibitem{Apolo:2020qjm}
L.~Apolo, H.~Jiang, W.~Song, and Y.~Zhong, ``{Modular Hamiltonians in flat
  holography and (W)AdS/WCFT},'' {\em JHEP} {\bf 09} (2020) 033,
  \href{http://www.arXiv.org/abs/2006.10741}{{\tt 2006.10741}}.

\bibitem{Chaturvedi:2018uov}
P.~Chaturvedi, Y.~Gu, W.~Song, and B.~Yu, ``{A note on the complex SYK model
  and warped CFTs},'' {\em JHEP} {\bf 12} (2018) 101,
  \href{http://www.arXiv.org/abs/1808.08062}{{\tt 1808.08062}}.

\bibitem{Apolo:2018oqv}
L.~Apolo, S.~He, W.~Song, J.~Xu, and J.~Zheng, ``{Entanglement and chaos in
  warped conformal field theories},'' {\em JHEP} {\bf 04} (2019) 009,
  \href{http://www.arXiv.org/abs/1812.10456}{{\tt 1812.10456}}.

\bibitem{Jensen:2017tnb}
K.~Jensen, ``{Locality and anomalies in warped conformal field theory},'' {\em
  JHEP} {\bf 12} (2017) 111, \href{http://www.arXiv.org/abs/1710.11626}{{\tt
  1710.11626}}.

\bibitem{Bhattacharyya:2022ren}
A.~Bhattacharyya, G.~Katoch, and S.~R. Roy, ``{Complexity of warped conformal
  field theory},'' \href{http://www.arXiv.org/abs/2202.09350}{{\tt
  2202.09350}}.

\bibitem{Aggarwal:2019iay}
A.~Aggarwal, A.~Castro, and S.~Detournay, ``{Warped Symmetries of the Kerr
  Black Hole},'' {\em JHEP} {\bf 01} (2020) 016,
  \href{http://www.arXiv.org/abs/1909.03137}{{\tt 1909.03137}}.

\bibitem{qiao2008edgestatesdestroyeddisordered}
Z.~H. Qiao, J.~Wang, Q.~F. Sun, and H.~Guo, ``How edge states are destroyed in
  disordered mesoscopic samples?,'' 2008.

\bibitem{Masina:2008zv}
I.~Masina and A.~Notari, ``{The Cold Spot as a Large Void: Rees-Sciama effect
  on CMB Power Spectrum and Bispectrum},'' {\em JCAP} {\bf 02} (2009) 019,
  \href{http://www.arXiv.org/abs/0808.1811}{{\tt 0808.1811}}.

\bibitem{Compere:2009zj}
G.~Compere and S.~Detournay, ``{Boundary conditions for spacelike and timelike
  warped $AdS_{3}$ spaces in topologically massive gravity},'' {\em JHEP} {\bf
  08} (2009) 092, \href{http://www.arXiv.org/abs/0906.1243}{{\tt 0906.1243}}.

\bibitem{Brown:1986nw}
J.~D. Brown and M.~Henneaux, ``{Central Charges in the Canonical Realization of
  Asymptotic Symmetries: An Example from Three-Dimensional Gravity},'' {\em
  Commun.Math.Phys.} {\bf 104} (1986)
207--226.

\bibitem{Barnich:2006av}
G.~Barnich and G.~Compere, ``{Classical central extension for asymptotic
  symmetries at null infinity in three spacetime dimensions},'' {\em
  Class.Quant.Grav.} {\bf 24} (2007) F15--F23,
\href{http://www.arXiv.org/abs/gr-qc/0610130}{{\tt gr-qc/0610130}}.

\bibitem{Strominger:1997eq}
A.~Strominger, ``{Black hole entropy from near horizon microstates},'' {\em
  JHEP} {\bf 02} (1998) 009,
  \href{http://www.arXiv.org/abs/hep-th/9712251}{{\tt hep-th/9712251}}.

\bibitem{Banados:1992wn}
M.~Banados, C.~Teitelboim, and J.~Zanelli, ``{The Black hole in
  three-dimensional space-time},'' {\em Phys. Rev. Lett.} {\bf 69} (1992)
  1849--1851,
\href{http://www.arXiv.org/abs/hep-th/9204099}{{\tt hep-th/9204099}}.

\bibitem{Banados:1992gq}
M.~Banados, M.~Henneaux, C.~Teitelboim, and J.~Zanelli, ``{Geometry of the
  (2+1) black hole},'' {\em Phys. Rev.} {\bf D48} (1993) 1506--1525,
  \href{http://www.arXiv.org/abs/gr-qc/9302012}{{\tt gr-qc/9302012}}.
[Erratum: Phys. Rev.D88,069902(2013)].

\bibitem{DESER2000409}
S.~Deser, R.~Jackiw, and S.~Templeton, ``Topologically massive gauge
  theories,'' {\em Annals of Physics} {\bf 281} (2000), no.~1, 409 -- 449.

\bibitem{Banados:2005da}
M.~Banados, G.~Barnich, G.~Compere, and A.~Gomberoff, ``{Three dimensional
  origin of Godel spacetimes and black holes},'' {\em Phys. Rev. D} {\bf 73}
  (2006) 044006, \href{http://www.arXiv.org/abs/hep-th/0512105}{{\tt
  hep-th/0512105}}.

\bibitem{Compere:2007in}
G.~Compere and S.~Detournay, ``{Centrally extended symmetry algebra of
  asymptotically Godel spacetimes},'' {\em JHEP} {\bf 03} (2007) 098,
  \href{http://www.arXiv.org/abs/hep-th/0701039}{{\tt hep-th/0701039}}.

\bibitem{Israel:2004vv}
D.~Israel, C.~Kounnas, D.~Orlando, and P.~M. Petropoulos, ``{Electric/magnetic
  deformations of S**3 and AdS(3), and geometric cosets},'' {\em Fortsch.
  Phys.} {\bf 53} (2005) 73--104,
  \href{http://www.arXiv.org/abs/hep-th/0405213}{{\tt hep-th/0405213}}.

\bibitem{Hofman:2014loa}
D.~M. Hofman and B.~Rollier, ``{Warped Conformal Field Theory as Lower Spin
  Gravity},'' {\em Nucl. Phys. B} {\bf 897} (2015) 1--38,
  \href{http://www.arXiv.org/abs/1411.0672}{{\tt 1411.0672}}.

\bibitem{Witten:1988hc}
E.~Witten, ``{(2+1)-Dimensional Gravity as an Exactly Soluble System},'' {\em
  Nucl.Phys.} {\bf B311} (1988)
46.

\bibitem{Henneaux:2019sjx}
M.~Henneaux, W.~Merbis, and A.~Ranjbar, ``{Asymptotic dynamics of AdS$_3$
  gravity with two asymptotic regions},''
\href{http://www.arXiv.org/abs/1912.09465}{{\tt 1912.09465}}.

\bibitem{Merbis:2019wgk}
W.~Merbis and M.~Riegler, ``{Geometric actions and flat space holography},''
  {\em JHEP} {\bf 02} (2020) 125,
  \href{http://www.arXiv.org/abs/1912.08207}{{\tt 1912.08207}}.

\bibitem{Cornalba:2002fi}
L.~Cornalba and M.~S. Costa, ``{A New cosmological scenario in string
  theory},'' {\em Phys. Rev. D} {\bf 66} (2002) 066001,
  \href{http://www.arXiv.org/abs/hep-th/0203031}{{\tt hep-th/0203031}}.

\bibitem{Cornalba:2002nv}
L.~Cornalba, M.~S. Costa, and C.~Kounnas, ``{A Resolution of the cosmological
  singularity with orientifolds},'' {\em Nucl. Phys. B} {\bf 637} (2002)
  378--394, \href{http://www.arXiv.org/abs/hep-th/0204261}{{\tt
  hep-th/0204261}}.

\bibitem{Azeyanagi:2018har}
T.~Azeyanagi, S.~Detournay, and M.~Riegler, ``{Warped Black Holes in Lower-Spin
  Gravity},'' {\em Phys. Rev. D} {\bf 99} (2019), no.~2, 026013,
  \href{http://www.arXiv.org/abs/1801.07263}{{\tt 1801.07263}}.

\bibitem{Afshar:2019tvp}
H.~R. Afshar, ``{Warped Schwarzian theory},'' {\em JHEP} {\bf 02} (2020) 126,
  \href{http://www.arXiv.org/abs/1908.08089}{{\tt 1908.08089}}.

\bibitem{Cotler:2018zff}
J.~Cotler and K.~Jensen, ``{A theory of reparameterizations for AdS$_3$
  gravity},'' {\em JHEP} {\bf 02} (2019) 079,
  \href{http://www.arXiv.org/abs/1808.03263}{{\tt 1808.03263}}.

\bibitem{Compere:2013bya}
G.~Comp\`ere, W.~Song, and A.~Strominger, ``{New Boundary Conditions for
  AdS3},'' {\em JHEP} {\bf 05} (2013) 152,
  \href{http://www.arXiv.org/abs/1303.2662}{{\tt 1303.2662}}.

\bibitem{Afshar:2015wjm}
H.~Afshar, S.~Detournay, D.~Grumiller, and B.~Oblak, ``{Near-Horizon Geometry
  and Warped Conformal Symmetry},'' {\em JHEP} {\bf 03} (2016) 187,
  \href{http://www.arXiv.org/abs/1512.08233}{{\tt 1512.08233}}.

\bibitem{Unterberger:2012}
J.~Unterberger and C.~Roger, {\em {The Schrödinger-Virasoro Algebra:
  Mathematical structure and dynamical Schrödinger symmetries}}.
\newblock Springer Berlin, 2012.

\bibitem{Guieu:2007}
L.~Guieu and C.~Roger, {\em L'Algèbre et le Groupe de Virasoro. : Aspects
  géométriques et algébriques, généralisations.(avec un appendice de Vlad
  Sergiescu)}.
\newblock 01, 2007.

\bibitem{Barnich:2017jgw}
G.~Barnich, H.~A. Gonzalez, and P.~Salgado-Rebolledo, ``{Geometric actions for
  three-dimensional gravity},'' {\em Class. Quant. Grav.} {\bf 35} (2018),
  no.~1, 014003, \href{http://www.arXiv.org/abs/1707.08887}{{\tt 1707.08887}}.

\bibitem{ELITZUR1989108}
S.~Elitzur, G.~Moore, A.~Schwimmer, and N.~Seiberg, ``Remarks on the canonical
  quantization of the chern-simons-witten theory,'' {\em Nuclear Physics B}
  {\bf 326} (1989), no.~1, 108--134.

\bibitem{DeKerf1997:Chap18}
E.~{De Kerf}, G.~Bäuerle, and A.~{Ten Kroode}, ``Chapter 18 extensions of lie
  algebras,'' in {\em Lie Algebras}, vol.~7 of {\em Studies in Mathematical
  Physics}, pp.~5--48.
\newblock North-Holland, 1997.

\bibitem{Maloney:2007ud}
A.~Maloney and E.~Witten, ``{Quantum Gravity Partition Functions in Three
  Dimensions},'' {\em JHEP} {\bf 02} (2010) 029,
  \href{http://www.arXiv.org/abs/0712.0155}{{\tt 0712.0155}}.

\bibitem{Giombi:2008vd}
S.~Giombi, A.~Maloney, and X.~Yin, ``{One-loop Partition Functions of 3D
  Gravity},'' {\em JHEP} {\bf 08} (2008) 007,
  \href{http://www.arXiv.org/abs/0804.1773}{{\tt 0804.1773}}.

\bibitem{Maloney:2009ck}
A.~Maloney, W.~Song, and A.~Strominger, ``{Chiral Gravity, Log Gravity and
  Extremal CFT},'' {\em Phys.Rev.} {\bf D81} (2010) 064007,
\href{http://www.arXiv.org/abs/0903.4573}{{\tt 0903.4573}}.

\bibitem{Anninos:2008fx}
D.~Anninos, W.~Li, M.~Padi, W.~Song, and A.~Strominger, ``{Warped AdS(3) Black
  Holes},'' {\em JHEP} {\bf 0903} (2009) 130,
\href{http://www.arXiv.org/abs/0807.3040}{{\tt 0807.3040}}.

\bibitem{Barnich:2015mui}
G.~Barnich, H.~A. Gonzalez, A.~Maloney, and B.~Oblak, ``{One-loop partition
  function of three-dimensional flat gravity},'' {\em JHEP} {\bf 04} (2015)
  178, \href{http://www.arXiv.org/abs/1502.06185}{{\tt 1502.06185}}.

\bibitem{Castro:2017mfj}
A.~Castro, C.~Keeler, and P.~Szepietowski, ``{Tweaking one-loop determinants in
  AdS$_{3}$},'' {\em JHEP} {\bf 10} (2017) 070,
  \href{http://www.arXiv.org/abs/1707.06245}{{\tt 1707.06245}}.

\bibitem{Denef:2009kn}
F.~Denef, S.~A. Hartnoll, and S.~Sachdev, ``{Black hole determinants and
  quasinormal modes},'' {\em Class. Quant. Grav.} {\bf 27} (2010) 125001,
  \href{http://www.arXiv.org/abs/0908.2657}{{\tt 0908.2657}}.

\bibitem{Song:2016gtd}
W.~Song, Q.~Wen, and J.~Xu, ``{Modifications to Holographic Entanglement
  Entropy in Warped CFT},'' {\em JHEP} {\bf 02} (2017) 067,
  \href{http://www.arXiv.org/abs/1610.00727}{{\tt 1610.00727}}.

\bibitem{Carlip:1994gc}
S.~Carlip and C.~Teitelboim, ``{Aspects of black hole quantum mechanics and
  thermodynamics in (2+1)-dimensions},'' {\em Phys. Rev. D} {\bf 51} (1995)
  622--631, \href{http://www.arXiv.org/abs/gr-qc/9405070}{{\tt gr-qc/9405070}}.

\bibitem{Maldacena:1998bw}
J.~M. Maldacena and A.~Strominger, ``{AdS(3) black holes and a stringy
  exclusion principle},'' {\em JHEP} {\bf 12} (1998) 005,
  \href{http://www.arXiv.org/abs/hep-th/9804085}{{\tt hep-th/9804085}}.

\bibitem{Aggarwal:2020igb}
A.~Aggarwal, L.~Ciambelli, S.~Detournay, and A.~Somerhausen, ``{Boundary
  conditions for warped AdS$_{3}$ in quadratic ensemble},'' {\em JHEP} {\bf 22}
  (2020) 013, \href{http://www.arXiv.org/abs/2112.13116}{{\tt 2112.13116}}.

\bibitem{Guichard}
D.~Guichard, {\em An Introduction to Combinatories and Graph Theory}.
\newblock Department of Mathematics Whitman College, 2023.

\end{thebibliography}\endgroup
\bibliographystyle{fullsort}
\end{document}